\documentclass[%
 aip,
 amsmath,amssymb,
 reprint,%
]{revtex4-1}

\usepackage{graphicx}% Include figure files
\usepackage{dcolumn}% Align table columns on decimal point
\usepackage{bm}% bold math
\usepackage[utf8]{inputenc}
\usepackage[T1]{fontenc}
\usepackage{mathptmx}
\usepackage{etoolbox}
\usepackage{gensymb}
\usepackage{url}
\usepackage{soul}
\usepackage[version=3]{mhchem}
\usepackage[usenames,dvipsnames]{xcolor}

\makeatletter
\def\@email#1#2{%
 \endgroup
 \patchcmd{\titleblock@produce}
  {\frontmatter@RRAPformat}
  {\frontmatter@RRAPformat{\produce@RRAP{*#1\href{mailto:#2}{#2}}}\frontmatter@RRAPformat}
  {}{}
}%
\makeatother
\begin{document}

\preprint{AIP/123-QED}

\title[Assessing the persistence of chalcogen bonds in solution with neural network potentials]{Assessing the persistence of chalcogen bonds in solution\\ with neural network potentials}

\author{V. Jur\'{a}skov\'{a}}
 \affiliation{Laboratory for Computational Molecular Design (LCMD), Institute of Chemical Sciences and Engineering, Ecole Polytechnique F\'{e}d\'{e}rale de Lausanne (EPFL), Lausanne, 1015, Switzerland.}
 
\author{F. C\'{e}lerse}
\affiliation{Laboratory for Computational Molecular Design (LCMD), Institute of Chemical Sciences and Engineering, Ecole Polytechnique F\'{e}d\'{e}rale de Lausanne (EPFL), Lausanne, 1015, Switzerland.}

\author{R. Laplaza}
\affiliation{Laboratory for Computational Molecular Design (LCMD), Institute of Chemical Sciences and Engineering, Ecole Polytechnique F\'{e}d\'{e}rale de Lausanne (EPFL), Lausanne, 1015, Switzerland.}
\affiliation{National Center for Competence in Research-Catalysis (NCCR-Catalysis), \'Ecole Polytechnique F\'ed\'erale de Lausanne, 1015 Lausanne, Switzerland.}

\author{C. Corminboeuf}
\affiliation{Laboratory for Computational Molecular Design (LCMD), Institute of Chemical Sciences and Engineering, Ecole Polytechnique F\'{e}d\'{e}rale de Lausanne (EPFL), Lausanne, 1015, Switzerland.}
\affiliation{National Center for Competence in Research-Catalysis (NCCR-Catalysis), \'Ecole Polytechnique F\'ed\'erale de Lausanne, 1015 Lausanne, Switzerland.}
\affiliation{National Centre for Computational Design and Discovery of Novel Materials
(MARVEL), \'Ecole Polytechnique F\'ed\'erale de Lausanne, 1015 Lausanne, Switzerland.}
\email{clemence.corminboeuf@epfl.ch}

\date{\today}% It is always \today, today,

\begin{abstract}

Non-covalent bonding patterns are commonly harvested as a design principle in the field of catalysis, supramolecular chemistry and functional materials to name a few. Yet, their computational description generally neglects finite temperature and environment effects, which promote competing interactions and alter their static gas-phase properties. Recently, neural network potentials (NNPs) trained on Density Functional Theory (DFT) data have become increasingly popular to simulate molecular phenomena in condensed phase with an accuracy comparable to \textit{ab initio} methods. To date, most applications have centered on solid-state materials or fairly simple molecules made of a limited number of elements. Herein, we focus on the persistence and strength of chalcogen bonds involving a benzotelluradiazole in condensed phase. While the tellurium-containing heteroaromatic molecules are known to exhibit pronounced interactions with anions and lone pairs of different atoms, the relevance of competing intermolecular interactions, notably with the solvent, are complicated to monitor experimentally but also challenging to model at an accurate electronic structure level.  
Here, we train a direct and baselined NNPs to reproduce hybrid DFT energies and forces in order to identify what are the most prevalent non-covalent interactions occurring in a solute-\ce{Cl-}-THF mixture. 
The simulations in explicit solvent highlight the clear competition with chalcogen bonds formed with the solvent and the short-range directionality of the interaction with direct consequences for the molecular properties in the solution. The comparison with other potentials (\textit{e.g.}, AMOEBA, direct NNP and continuum solvent model) also demonstrates that baselined NNPs offer a reliable picture of the non--covalent interaction interplay occurring in solution.
\end{abstract}

\maketitle

\section{Introduction}

Non--covalent interactions (NCIs) impact all areas of chemistry. Among many examples, they govern the structural stability and activity of proteins and DNA, \cite{frieden1975non,vcerny2007non} influence the regioselectivity and enantioselectivity in organocatalytic reactions, \cite{davis2017harnessing,orlandi2017parametrization,proctor2020exploiting,zuend2009mechanism,uyeda2011transition} and are used as a design principle to build functional materials.\cite{ho2020chalcogen} Notable NCIs include hydrogen bonds, $\pi$--$\pi$ interactions, anion/cation--$\pi$ interactions and $\sigma$-hole interactions. While textbook examples commonly depict non-covalently bonded patterns using a static dimer picture, realistic solvated chemical environments involve molecules undergoing a large number of competing NCIs. Favored conformations then arise from a subtle equilibrium between entropic and enthalpic contributions relative to all the possible NCIs. Yet, identifying which interaction plays the most prominent role is not always trivial and even less when solvent molecules intervene.

Achieving an accurate description of competing NCIs in a condensed environment, and evaluating their relative importance has thus been a long-lasting goal of computational chemistry. \cite{riley2011noncovalent,meyer2019dori,hobza2006world,maharramov2016non} Traditional static computations in implicit solvent provide an accurate estimation of the strength of individual NCIs, but at the cost of neglecting the role of thermal fluctuations, full entropic effects and detailed descriptions of the solvation mechanism. \cite{schneider2019quantification}

The joint experimental and computational investigations by Cockroft and coworkers further stress the difficulties of reconciling theory and experiment when taking the solvent into consideration.\cite{Elmi2021,mati2013}  
They have, for instance, demonstrated that NCIs are strongly attenuated in a solvated environment.\cite{Yang2013} Despite the intense efforts associated with developing \textit{a posteriori} dispersion corrections\cite{Becke2005,Becke2005a,Johnson2005,Becke2006,Becke2007,Grimme2010a,Grimme2011wires,Grimme2011jcomputchem,Grimme2016,Tkatchenko2009,Tkatchenko2012,Steinmann2010,Steinmann2011,Steinmann2011jchemphys,Corminboeuf2014,ehrlich2011system,Caldeweyher2019,Caldeweyher2020} or with reducing the cost of wavefunction-based methods,\cite{Neese2009,DePrince2013_1,DePrince2013_2,Riplinger2013,Parrish2014,Sparta2014,Fales2020} the finite-temperature description of competing NCIs remains coarse if the solvent is represented as a continuum. In particular, interaction energies estimated by dispersion-corrected computations in implicitly modeled solvent can be one magnitude larger than experimental energies in solutions.\cite{Yang2013,Pollice2017,Pollice2019,Hansen2014} While the gas-phase data are reproduced accurately by the underlying dispersion-corrected methods, the observed discrepancy suggests the inadequate description of the dispersion compensation and solute-solvent interactions by current implicit solvent models.\cite{Pollice2017,Pollice2019} As the influence of solvation on the dispersion forces is still being actively investigated,\cite{Schumann2021} proper care should be placed on the unambiguous quantification and characterization of NCIs in solution when implicit solvent is used.\cite{Grimme2012,Schneider2019} 

To overcome the shortcomings associated with the simplified picture provided by implicit solvent models and achieve a thorough statistically converged dynamic representation of all the relevant driving forces behind the formation of minima stabilized by various NCIs, it is necessary to rely upon finite temperature molecular dynamic (MD) simulations explicitly including the solvent molecules. Yet, the description of competing non-covalent interactions is a challenge for both classical and \textit{ab initio} MD, although for different reasons. Classical force fields, for instance, suffer from the inaccurate description of the electrostatic contributions beyond the monopole approximation. While polarizable force fields (\textit{e.g.},  AMOEBA,\cite{ponder2010current} SIBFA, \cite{devereux2014supervised} NEMO, \cite{holt2010nemo} GEM \cite{torabifard2015development}) aim at addressing this issue, their main focus currently remains on the description of proteins and metalloproteins with little emphasis placed on organic molecules or on systems containing somewhat exotic elements. Alternatively, QM/MM or full \textit{ab initio} MD provide an accurate description of individual NCIs, but simulating the free-energy landscapes of non-covalent patterns in a solvated environment is limited by the insufficiently long time scales. The challenge associated with achieving statistically converged simulations at quantum chemical accuracy thus remains present.

Stemming from the advances in Neural Networks,\cite{balabin2009neural} high dimensional neural network-based potentials (NNPs)\cite{behler2007generalized,behler2017first} trained on DFT or post-Hartree-Fock energies (and possibly forces), as pioneered by Behler and Parrinello, provide an efficient and accurate alternative to traditional \textit{ab initio} MD. These approaches are increasingly used in the simulation of solid-state systems,\cite{PhysRevB.83.153101,PhysRevMaterials.5.063804} molecules in the gas phase,\cite{} liquids\cite{Morawietz8368,Cheng2020} or molecules on the interfaces. \cite{C6CP05711J} Their application in the simulations of molecular systems in solution however remains scarce.\cite{yang2021using,Schran2021} 
Recently, some of us demonstrated the advantages of applying NNPs to go beyond the semi-empirical description of reactions involving hydrogen peroxide, acid and phenol in an amphiphilic environment.\cite{rossi2020simulating} The neural network-based simulations showed the importance of relying upon accurate potentials and achieving long timescale for properly capturing the dynamic of the protonation states of the strong acid and the competition between different hydrogen bonds.    

\begin{figure*}[!t]
    \centering
    \includegraphics[scale=0.4]{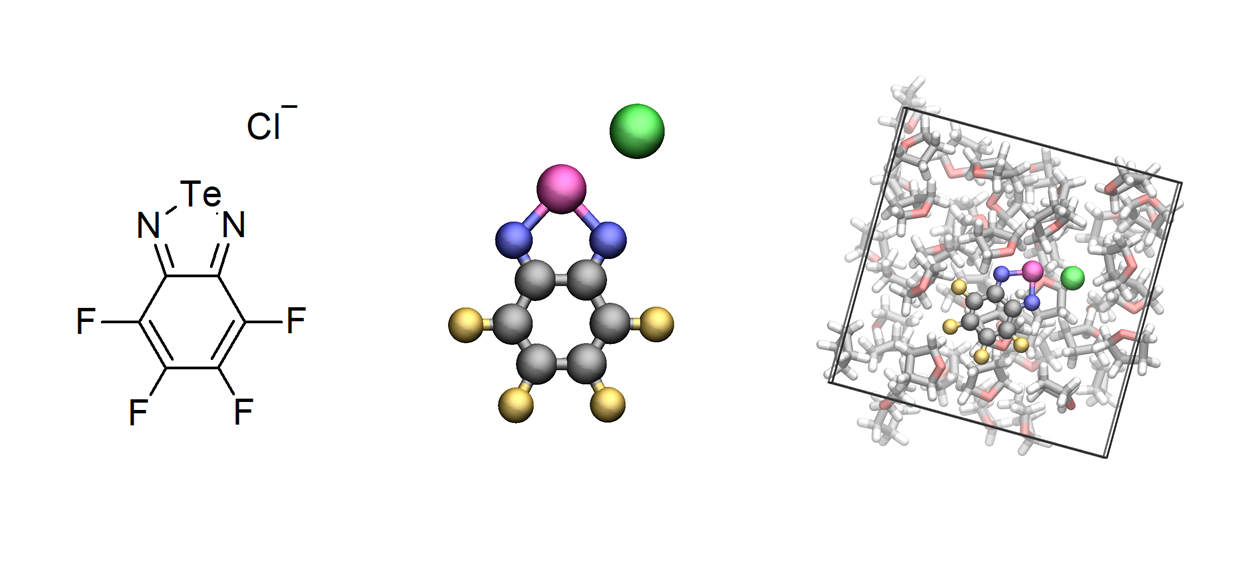}
    \caption{Chalcogen bond between the benzotelluradiazole and \ce{Cl-}. Left and middle: the chemical structure of the bonded complex, right: solvated system used in the simulations - H, C, N, O, F, Cl and Te atoms are depicted in white, gray, blue, red, yellow, green, and magenta, respectively. }
    \label{benzo-depict}
\end{figure*} 

Here, we target the description of competing NCIs in solution. Specifically, we focus on chalcogen bonds as a relatively unexplored interaction\cite{aakeroy2019definition} with increasing applicability for the design of organocatalyts, anion transport through lipidic membranes, supramolecular architectures and functional materials. \cite{https://doi.org/10.1002/anie.201910639,doi:10.1021/jacs.6b05779,macchione2018mechanosensitive,mahmudov2017chalcogen} The existence of chalcogen bonds is commonly understood as a result of the anisotropic charge density distribution around a covalently bound main group atom, which leads to electron-deficient $\sigma$-hole regions that interact with electron-rich moieties. Chalcogen bonds are non-negligible in strength and of marked importance, as common chalcogen-containing molecules are interacting through their $\sigma$-holes with lone pairs on \ce{N} and \ce{O} atoms and other prevalent Lewis bases.

A necessary prerequisite for warranting the relevance of chalcogen bonding in, for instance, molecular recognition or catalysis is to ensure strong self-association.
A relevant heteroaromatic compound inclined to form strong chalcogen bonds in solution through N···chalcogen interactions is 4,5,6,7--tetrafluorobenzo-2,1,3--telluradiazole (further referred only as benzotelluradiazole, see Fig. \ref{benzo-depict}) and \ce{Cl-} anion. The association constant for this complex in tetrahydrofuran (THF) was recently measured by means of UV-vis absorbance spectroscopy.\cite{garrett2015chalcogen} The free energy corresponding to this interaction was estimated to be 7.0 kcal/mol, which aligns well with the chalcogen bond computed at the B97-D3/def2-TZVP level with THF represented by Polarizable Continuum Model (PCM).\cite{garrett2015chalcogen} Apart from the interaction between tellurium and the \ce{Cl-} anion, benzotelluradiazole can in principle promote alternative anion-$\pi$ interactions as well as those involving the lone pair of the solvent molecules. 

To provide a realistic and atomistic dynamical description of chalcogen bonds in the condensed phase, we exploit baselined and direct NNPs similar to previous work.\cite{rossi2020simulating} The choice of the reliable reference method for the baselined NNPs is essential to ensure the description of a broad range of NCIs. As seen in the recent benchmark study by Řezáč et al.\cite{Rezac2020,Rezac2020a,Kriz2021} and Goerigk and coworkers,\cite{Mehta2021} the characterization of the interaction strength of NCIs and chalcogen bonding, in particular, is problematic for most popular DFT approximations owing to the possible role the density errors and its worsening by dispersion corrected schemes. While the most reliable characterization is achieved by modern double hybrid DFT functionals, hybrid DFT functionals were generally shown to be a reasonable compromise.\cite{Mehta2021} Here we choose PBE0\cite{adamo1999toward}-D3BJ\cite{Grimme2011jcomputchem} as the reference level. We also interface the i-PI dynamic driver with the xTB code to enhance the versatility of the baselined NNPs approach and enable the simulations of all the chemical elements present in the investigated system.

The objective of the simulation workflow discussed below is to address the following compelling questions regarding the nature and behavior of chalcogen bonds in solution: How strong and dominant is the interaction in tetrahydrofuran? Is the chalcogen bond and associated $\sigma$-hole responsible for the observed association between \ce{Cl-} and benzotelluradiazole or is it a consequence of more complicated interplay between the NCIs? Is the directionality of the chalcogen bond preserved in solution? How well are the relevant solvation effects captured by implicit solvent models?

These questions are addressed through careful comparison of molecular dynamics simulations obtained from two NNPs approaches, the AMOEBA polarizable force field, and static implicit solvent computations.

\section{Methods and computational details}

\begin{figure*}[!t]
    \centering
    \includegraphics[scale=0.9]{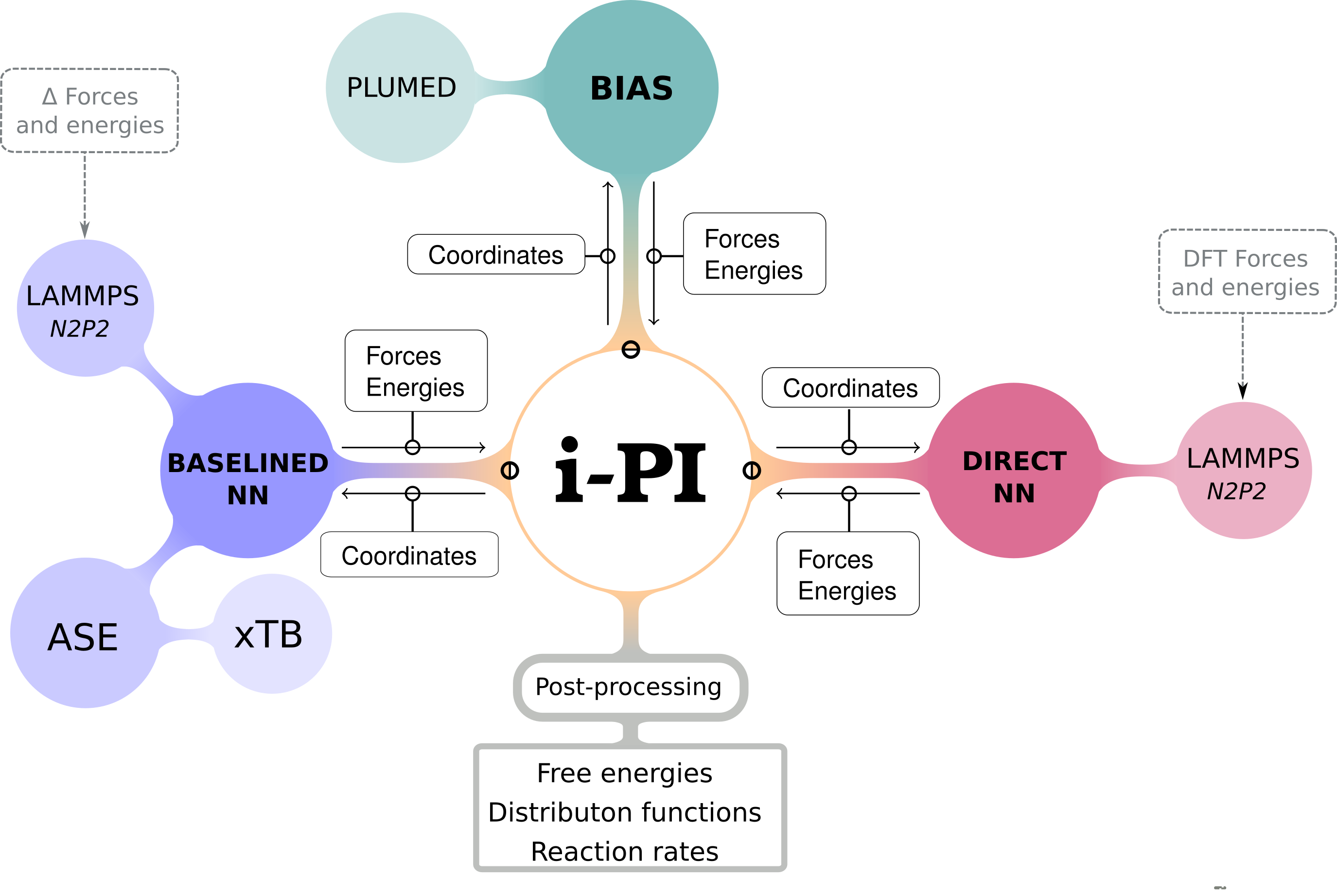}
    \caption{Scheme of a simulation workflow and software packages utilized in this work.}
    \label{workflow}
\end{figure*} 

\paragraph{Overview}

The workflow used herein (Fig. \ref{workflow}) is a variant of the baselined (BNNP) and direct (DNNP) NNPs training procedure published previously.\cite{rossi2020simulating} The procedure begins with constructing a robust database of structures covering the relevant part of the conformational space (\textit{i.e.}, chalcogen bond and anion-$\pi$ interactions), determined from umbrella sampling and temperature replica-exchange simulations at a baseline electronic structure level, which here corresponds to a GFN0-xTB\cite{https://doi.org/10.1002/wcms.1493} Hamiltonian. Energies and forces are then recomputed at a more accurate level (PBE0\cite{adamo1999toward} with D3BJ correction\cite{Grimme2011jcomputchem}) that serve as reference data. Behler--Parrinello NNPs\cite{behler2007generalized,singraber2019parallel} are then trained to reproduce the reference data either directly or using a difference between the baseline and reference method.\cite{ramakrishnan2015big} Molecular dynamics together with enhanced sampling techniques are run with each NNPs or in a combined multiple time step (MTS) scheme.\cite{tuck+92jcp}

Details on the simulation protocol, preparation and selection of the structures and symmetry functions chosen for the NNPs training and molecular dynamics trajectory are provided in the following sections. A technical description of the electronic structure methods and force field parametrization employed as a comparison is provided in the Supplementary Material (SM).

\paragraph{Baselined NNPs using the GFN0-xTB semiempirical potential}

The superior accuracy and stability of kernel-based models and NNPs, trained from a baselined electronic structure level (\textit{i.e.}, $\Delta$-ML\cite{ramakrishnan2015big}) has been largely demonstrated.\cite{Shen2016,Bartok2017,Meyer2018,Sun2019,Fabregat2020jctc,rossi2020simulating,Bogojeski2020}

This work uses GFN0--xTB as a baseline potential as the approach is well-suited for simulating solvated systems featuring a large diversity of chemical elements and environments. The GFN\textit{n}--xTB family of Hamiltonians \cite{https://doi.org/10.1002/wcms.1493} goes beyond the limits of pairwise parametrization imposed by several semiempirical approaches, such as the related DFTB. \cite{doi:10.1021/ct300849w} In particular, the element-specific empirical fitting enables derivation of parameters that cover the majority of the periodic table (up to Z=86), making GFN\textit{n}--xTB suitable for computing the broad range of compounds and elements necessary here. GFN0--xTB, which is a non-self-consistent variant of GFN\textit{n}--xTB, is also significantly faster than other semiempirical approaches. 

To integrate GFN0-xTB into the working simulation machinery, we provide a calculator linking the xTB code and Atomistic simulation environment (ASE)\cite{larsen2017atomic}. The main advantage in calling xTB via ASE is the existence of a socket interface between ASE and i-PI, which share the information between xTB and ASE without direct modifiction of the codes. The ASE calculator for the xTB version 6.2 is available on Github (\url{https://github.com/lcmd-epfl/xtb_ase_io_calculator}).

\paragraph{Training set construction}

The system is composed of one benzotelluradiazole solvated in 45 tetrahydrofuran (THF) molecules and one \ce{Cl-} anion, yielding 599 atoms in a 18.428 \AA~cubic box (see Fig. \ref{benzo-depict}). The box was minimized and equilibrated to the experimental THF density using the General Amber Force Field in the sander module from AmberTools 16.\cite{case2005amber} The solute molecules were kept frozen in the geometry optimized at the M06-2X/def2-TZVP level during the equilibration run. Steered Molecular Dynamics\cite{izrailev1999steered} were used to generate structures with a Te--Cl distance ranging from 2.5 to 7 \AA. 19 structures from the Steered MD runs were then used to perform umbrella sampling (US)\cite{torrie1977nonphysical} simulations using the GFN0--xTB potential. Each window was simulated for 50 ps using the Abin code.\cite{daniel_hollas_2018_1228463} The training set was constructed from 2 000 structures selected from the US simulation with Farthest Point Sampling (FPS).\cite{imbalzano2018automatic} Additional sets of 300 and 1500 structures respectively were generated by performing high temperature (1000 K) and two Replica Exchange MD (REMD)\cite{sugita1999replica} simulations within the 273.15 - 700 K temperature range. The first REMD simulation was initiated from structures featuring a chalcogen--bonded \ce{Cl^-}. To assess the relevance of the anion-$\pi$ interaction between \ce{Cl^-} and  benzotelluradiazole, the second REMD simulation was initiated with \ce{Cl^-} placed above the  benzotelluradiazole plane. Following a first round of NN training (\textit{vide infra}), the training set was enlarged with 1200 additional structures taken from a trial US simulations with the trained BNNP. The complete set thus contained 5 000 structures with their respective DFT energies and forces (2 000 from GFN0--xTB US, 1 800 from GFN0--xTB REMD, 1 200 from US using the BNNP trained with the previous 3 800 structures). Such a set ensured configurational diversity and sufficient coverage of all the relevant regions of the potential energy landscape.

\paragraph{Atomic Symmetry Functions (ASFs) definition}

We used radial ($G_2$) and angular ($G_4$) Behler--Parrinello atomic symmetry functions (ASFs) \cite{behler2007generalized,behler2011atom} as descriptors in the NNPs. ASFs were built for each elements using a cutoff equal to 16 a.u. for BNNP and 14 a.u. for DNNP. We initially generated 273 radial and 2064 angular ASFs per element for the BNNP and 784 and 2064 radial and angular ASFs for the DNNP. The radial functions for the DNNP were generated on a denser grid to provide a better resolution and to stabilize the potential. 
We used CUR decomposition as described in Ref. \citenum{imbalzano2018automatic} to select the most relevant fingerprints for the system.  
Previous studies\cite{imbalzano2018automatic,rossi2020simulating} demonstrated that 64 symmetry functions per element yield sufficiently low training errors for systems containing 1 to 4 different elements. Regarding the larger number of species in the mixture, we monitored the root-mean-square error (RMSE) in the training of energies and forces for sets containing 64 to 256 symmetry functions per element. The results are summarized in the SM (see Fig. S2). Based on the training performance, we selected 128 ASFs per element for the BNNP and 256 ASFs per element for the DNNP, yielding 896 (resp. 1792) symmetry functions for the whole system.

\paragraph{NNPs training}

The DNNP and BNNP were trained using the Behler--Parrinello neural network potential architecture implemented in the n2p2 version 2.0.3.\cite{singraber2019parallel} The input values to the DNNP are PBE0-D3BJ energies and forces computed for every structure in the training set. The BNNP is trained to reproduce the difference between the DFT and GFN0-xTB values and therefore served as a correction applied to the semiempirical computations. All NNs contained 2 hidden layers and 22 nodes per layer. The training set structures were normalized before the training, and the symmetry functions were centered and scaled by standard deviation. \cite{singraber2019parallel} The complete training data set of 5 000 structures was split in 80\% for training and 20\% for testing. The weights were optimized using a Kalman filter. We use 100\% of the energies and 0.5\% of the force components per configuration in the training. An ensemble of 3 neural networks trained on different subsets of the training set was used to compute the BNNP energy and forces, and a single neural network was used for DNNP computations.

\paragraph{Molecular dynamics}

All the MD trajectories were propagated using the i--PI code, which was connected with the xTB--ASE interface, LAMMPS drivers for forces and energy computations \cite{kapil2019pi,singraber2019library} and Plumed 2.6.2 \cite{plumed2019,TRIBELLO2014604}, which served to apply restraints for US. To decrease the cost of BNNP, a multiple--time--step (MTS) scheme combining DNNP and BNNP (MTS-NNP) was employed with the BAOAB integrator. \cite{leimkuhler2016efficient} The forces and energies were computed every 3 fs by the GNF0-xTB and BNNP (\textit{i.e.}, the outer time step) and every 0.5 fs by the DNNP (\textit{i.e}, inner time step). The simulations using only DNNP were integrated with the time step 1 fs using 1 NNP. Stability of the trajectories was improved using both a velocity rescaling thermostat with a frequency of 10 fs and a thermostat based on the generalized Langevin equation \cite{ceriotti2010colored} (GLE) with the same parameters as in the Ref. \citenum{rossi2020simulating}. The extrapolation regime of the NNPs during the MD was monitored by printing the extrapolation warnings at every step. The extrapolation warning was triggered when some of the symmetry functions reached value outside the interval on which it was trained. 

\paragraph{Umbrella sampling}

Potential of Mean Force (PMF) was used to evaluate the strength of the interaction between benzotelluradiazole and \ce{Cl-}. We applied umbrella sampling and selected the distance between the chloride ion and the Te atom as a reaction coordinate (RC). The RC was decomposed into 19 windows, ranging from 2.5 \AA ~ to 7 \AA~ with a width of 0.25 \AA. A harmonic restraint of 50 kcal/mol$\cdot$ \AA$^2$ was applied in all windows. To improve the overlap among the windows in DNNP simulations, additional windows were added at 3.50, 5.00 - 5.20 \AA~ with a restraint 100 kcal/mol$\cdot$ \AA$^2$. 
We ran molecular dynamics with a total simulation time of 200 ps per window for MTS-NNP and 500 ps per window using DNNP alone, corresponding to a total simulation time of 4.8 ns and 9.5 ns respectively. For the sake of comparison, we simultaneously ran the same simulation with AMOEBA polarizable force field (10 ns/per window, total 190 ns). The first $\sim$ 20 ps of simulation for each window were used for equilibration and were thus removed from analysis. The free energy profile was reconstructed using the Weighted Histogram Analysis Method (WHAM) as implemented by Grossfield\cite{grossfield2003wham}.
The convergence of the PMF was verified by running the US in the reverse direction (see Fig. S7--S10). US with AMOEBA and DNNP exactly overlapped in both directions demonstrating sufficient convergence of the simulations. Reverse MTS-NNP PMF was $\sim$1 kcal/mol lower than the forward. Longer sampling might be therefore needed to achieve perfect overlap, but the difference between PMF was in an acceptable span. 

\section{Results and Discussion}

The main purpose of this work is to build efficient and accurate NNPs to simulate complex mixtures encompassing a variety of atom type and competing NCIs such as the chalcogen bond investigated herein. From the chemical perspective, our goal is to unravel the origin of the attractive interaction between benzotelluradiazole and \ce{Cl-} solvated in THF, and to investigate the significance and structural properties of chalcogen bond in the presence of competing NCIs. We first evaluate the accuracy of the trained DNNP and BNNP and validate their robustness on a test set. We then analyze the outcome of umbrella sampling simulations with these potentials and estimate the strength of the interaction in THF prior to characterizing its nature. Finally, we describe the structural properties of the chalcogen bond in THF, and assess the role played by the solvent and by other NCIs through comparisons with simulations performed with the AMOEBA polarizable force field and with static computations in PCM. 

\subsection*{Validation of the NNPs}
\begin{figure*}[!htp]
    \centering
    \includegraphics[scale=0.7]{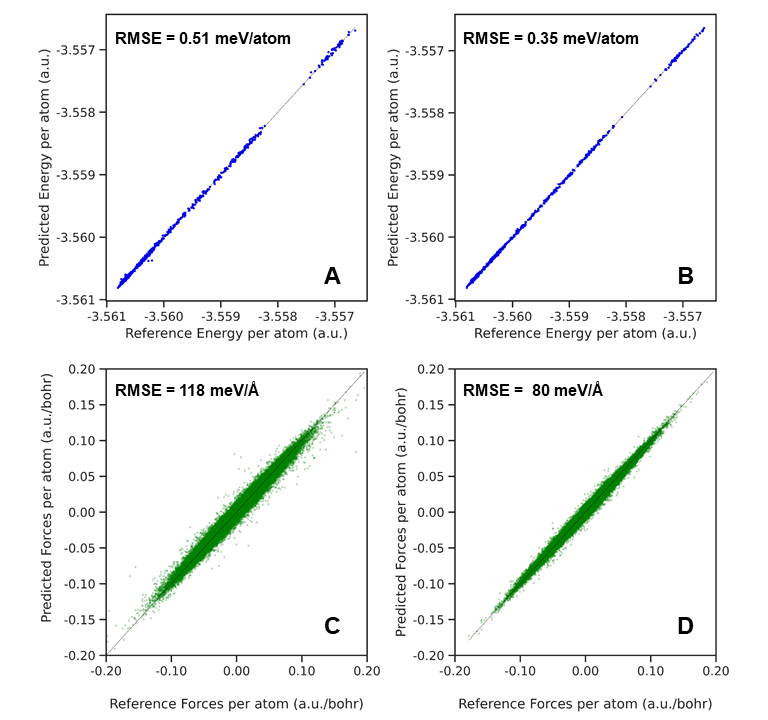}
    \caption{Parity plot for energy and forces of DNNP (A, C) and BNNP (B, D). Results are reported after 100 training epochs. Data are obtained for the 1000 structures contained in the test set.}
    \label{NN_parity_plot}
\end{figure*}
Fig. \ref{NN_parity_plot} depicts the parity plots (\textit{i.e.}, reference vs. predicted values) for BNNP and DNNP for the structures in the test set. Both NNPs achieved very low RMSE in the energy, \textit{i.e.}, $\sim$ 0.51 meV/atom for DNNP and $\sim$ 0.35 meV/atom for BNNP. The RMSE in the prediction of forces is equal to 118 meV/\AA/atom and 80 meV/\AA/atom for DNNP and BNNP, respectively.

To analyze the performance of the two NNPs variants and especially their ability to describe the variety of elements, the overall forces parity plots are split into individual plots for each element (see Fig. S4 -- S6 in SM). The reference and predicted forces show a good correlation, although larger errors are obtained for the elements in the solute, which are less represented than those in the solvent. The smallest error is obtained for the hydrogen atoms (48 meV/\AA~for BNNP and 85 meV/\AA~for DNNP) and the largest for tellurium (244 meV/\AA~for BNNP and 335 meV/\AA~for DNNP). Both the DNNP and BNNP forces display a better correlation with the reference DFT data than the original forces computed with GFN0-xTB, which are generally too low in absolute value. Similarly, both NNPs provide a very good description of the explicit solvent, as can be seen from very low errors in forces for the C, O and H atoms.

All methods yield stable trajectories during short non-biased MD simulations (\textit{i.e.}, 50 ps). DNNP however triggers slightly more extrapolation warnings than BNNP. Importantly, DNNP with the MTS set up does not extrapolate significantly while substantially decreasing the cost of the simulation. Following these promising results, we evaluate the stability and behavior of the DNNP and MTS-NNP in longer US simulations. 

The US simulations are more challenging for the NNPs, as the restraining potential can easily push the structures toward regions underrepresented in the training set. While both potentials yield stable dynamics at long timescales (\textit{i.e.}, more than hundreds of ps), the extrapolation warnings triggered in the US simulations are indeed larger than in non-biased runs for the DNNP simulations (see Figs. S12 -- S14). Alternatively, the MTS-NNP ensures lower extrapolation risk due to the robustness of the underlying BNNP. Given the relatively weak strength of non-covalent interactions, even a mild difference in the PES may significantly alter the observed interplay. The reliable prediction of energies and forces in the regions outside the training set, \textit{i.e.}, the ability of NNPs to generalize, is therefore crucial. To evaluate the impact of the extrapolation on the predictive power of the NNPs, we build an additional validation set containing 500 structures selected by FPS from the DNNP US trajectories and compute the reference DFT energy and forces. The structures are extracted from the regions with significant extrapolation warnings, \textit{i.e.}, regions with the Te-\ce{Cl-} distance above 3.5 \AA.

\begin{figure*}[!thp]
    \centering
    \includegraphics[scale=0.63]{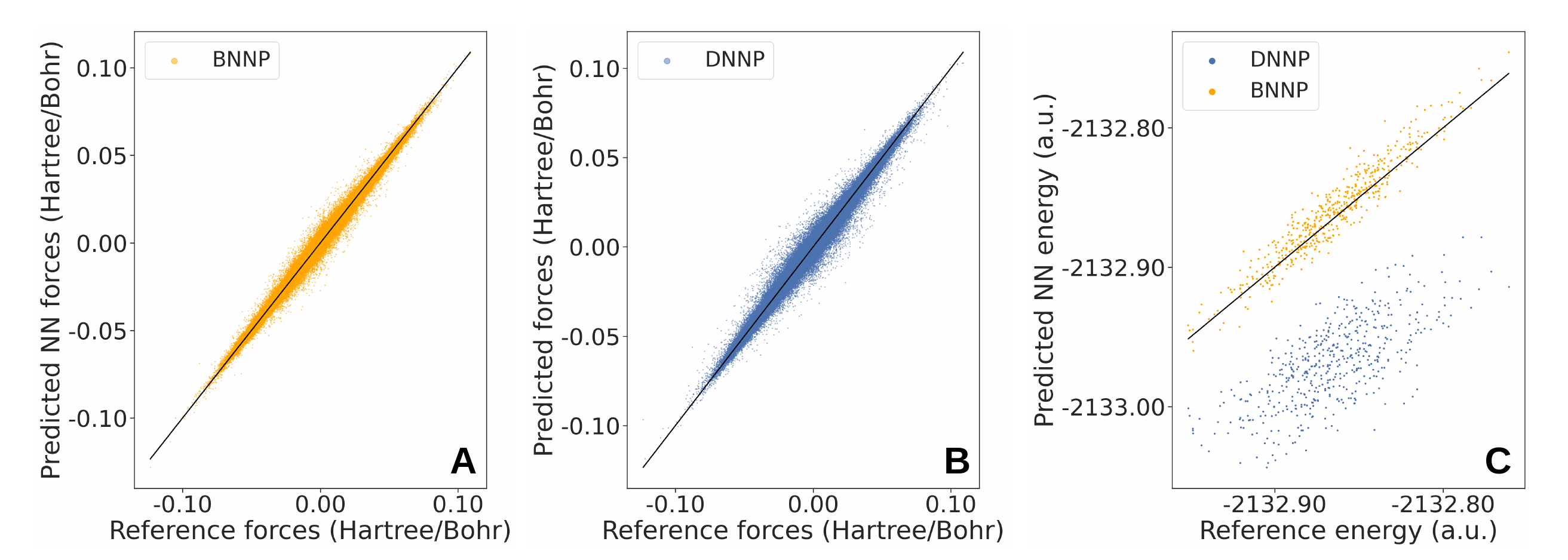}
    \caption{Energy (C) and forces (A, B) parity plots evaluated on an additional test set structures using both NNPs. RMSE for energy: 4.82 meV/atom (DNNP), 0.52 meV/atom (BNNP); RMSE for forces: 104 meV/\AA/atom and 72 meV/\AA/atom for DNNP and BNNP, respectively.}
    \label{validation}
\end{figure*}

The predictions of forces (Fig. \ref{validation}, A and B) for both NNPs are in a very good agreement with the PBE0-D3BJ reference, showing a RMSE of 72 meV/\AA/atom and 104 meV/\AA /atom for BNNP and DNNP, respectively. The main difference between the NNPs is however visible in the prediction of the energy (Fig. \ref{validation}, C). While the BNNP energy correlates well with the reference, and RMSE is consistent with the test set (RMSE 0.52 meV/atom), the DNNP strongly underestimates the energies of the selected out-of-sample structures (RMSE 4.82 meV/atom). The error in the energy predictions in these regions is therefore out of control. This behavior further highlights the need for sufficiently large training set and careful validation of the DNNP, as the apparent stability of the simulations can be misleading and cause misinterpretation of the results. These errors are especially relevant in the context of non-covalent interactions owing to their sensitivity to the quality of the PES. While the issue could be resolved by retraining of the neural networks including structures from the extrapolated regions, the perfect stabilization of the DNNP is out of the scope of this work. Importantly, MTS-NNP profits from the speed of the DNNP while preventing the extrapolation by corrections with more robust BNNP. Table \ref{tab:timings} demonstrates the computational gain resulting from the application of the MTS-NNP approach compared to the original hybrid DFT, GFN0-xTB, BNNP and DNNP computations. The MTS-NNP is more than $25       
 \times 10^4$ times faster than original DFT computations and 4 times faster than the BNNP alone. Both the GFN0-xTB and NN computations can also benefit from parallelization of the corresponding codes, which further speeds up the actual simulations.  
\begin{table}[tbh]
    \centering
    \begin{tabular}{ccccc}
         PBE0& GFN0-xTB & GFN0-xTB & MTS-NNP & DNNP \\
         -D3BJ & + BNNP comm. &  & comm. &  \\
         \hline
         \hline
         $ 227 \times 10^3$ & 37 & 29.7 & 8.8 &  2.5 \\
    \end{tabular}
\caption{CPU time (core seconds) required to perform MD with 0.5 fs time step. DFT timing is determined based on a single computing node with two 14 cores Intel Broadwell processors running at 2.6 GHz.
GFN0-xTB and NN timings are computed from single-core execution using a node with Intel Xeon processor running at 2.2 GHz. The BNNP comm. label indicates a prediction obtained with a committee of three BNNPs, DNNP indicates results from a single direct NN. The MTS cost is computed from a $\sim$ 53~CPU s timing for a  1-6 MTS scheme with 3~fs outer time step, consisting of 1 GFN0-xTB + 8 NN computations (3 BNNPs + 5 DNNPs).}
    \label{tab:timings}
\end{table}

The MTS-NNP is therefore perfectly suited for the long MD using enhanced sampling as it provides reliable generalization while significantly decreasing the computational cost. In the next section, we use MTS-NNP as a reference method to analyze the results of the US simulations, assess the impact of possible undesirable extrapolations in DNNP and evaluate performance of AMOEBA, GFN0-xTB and PCM in comparison with the NNPs.

\subsection*{Role of NCIs in benzotelluradiazole-\ce{Cl-} association} 
\begin{figure*}[!htp]
    \centering
    \includegraphics[scale=0.54]{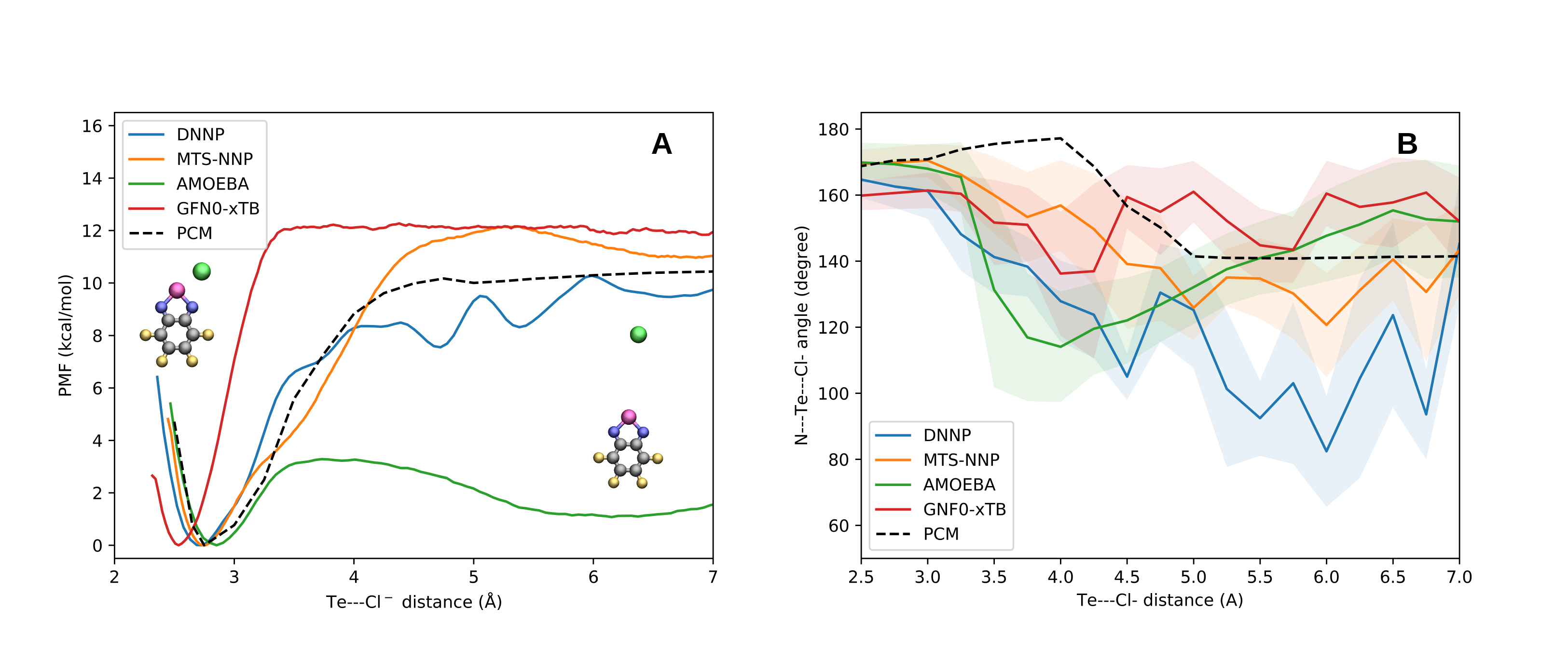}
    \caption{A: Comparison of PMF along the Te--\ce{Cl-} coordinate for DNNP, MTS-NNP, AMOEBA, and GFN0-xTB umbrella sampling simulations and relaxed scan in PCM. B: Evolution of the N--Te--Cl- angle along the collective variable. The average angle is computed for every window separately. Its standard deviation is represented with the shaded regions.}
    \label{PMF}
\end{figure*}

The strength of the interaction between benzotelluradiazole and \ce{Cl-} is evaluated \textit{via} the PMF (see Fig. \ref{PMF}A) reconstructed along the Te-\ce{Cl-} distance. The MTS-NNP predicts a favorable association of benzotelluradiazole and \ce{Cl-} with a free energy of 11 kcal/mol. The atomistic resolution of the free energy profiles reveals the directionality and nature of the bonding patterns associated with these energetics. The deep minimum located around the  2.75~\AA~ Te-\ce{Cl-} distance  coincides with a N-Te-\ce{Cl-} bond angle in the 160 - 180$\degree$ range (see Fig. \ref{PMF}, B). The short distance and nearly linear angle confirms the presence of a strong and directional interaction between benzotelluradiazole and \ce{Cl-}. The directionality does however persist only in the relatively short range, \textit{i.e.}, up to around 4.50 \AA~of Te-\ce{Cl-} distance. At the longer distances, the angle deviates from the liner arrangement and the mode of interaction changes. chalcogen bond is replaced by a different  The evolving interactions occurring in this region are modulated by the explicit solvation responsible for the energy maximum associated with the transition state between the benzotelluradiazole-\ce{Cl-} complex and the solvent-separated  structure (\textit{vide infra}).

 The PMF profiles modeled with other potentials (see Fig. \ref{PMF}A) substantially differ from the MTS-NNP one. The profile at the bare GFN0-xTB level is shifted towards lower Te-\ce{Cl-} distances. This behavior is likely caused by the wrong description of the repulsive and electrostatic part of the potential, a limitation which is well known as GFN0-xTB does not use self-consistent evaluation of the charges.\cite{https://doi.org/10.1002/wcms.1493} The difference between GFN0-xTB and MTS-NNP stresses the necessity of using a more accurate DFT level for achieving a proper description of the benzotelluradiazole-\ce{Cl-} interaction. 

DNNP, AMOEBA and the relaxed scan using a PCM all identify an energy minimum at a distance in agreement with MTS-NNP. Yet the free energy obtained with AMOEBA is dramatically lower (about 1.8 kcal/mol) than the 11 kcal/mol of MTS-NNP, which is itself slightly higher then the experimental value (\textit{i.e.}, 1.5 times stronger). Akin to related halogen bonds, charge transfer contributions are expected to be part of the interaction energy.\cite{wang2009chalcogen,Wolters2014,Wang2014,Rezac2017,Inscoe2021,Nunzi2021} While the exact role of charge transfer in $\sigma$--hole interactions is still under debate, \cite{thirman2018characterizing,huber2012unexpected,clark2007halogen} the  original  AMOEBA  force  field expression suffers from a lack of explicit charge transfer description,\cite{ponder2010current} which could explain its collapse in the long–range. An adjustment of the multipoles (including higher order terms) and Van der Waals terms\cite{wu2012automation} (by softening the VdW radii\cite{demerdash2017assessing,das2018improvements}) could be a route to retrieve the missing charge transfer term. Similarly to MTS-NNP, the AMOEBA profile displays a maximum coinciding with the transition state, but at a shorter distance, as expected from the weaker interaction energy.

The ~10 kcal/mol free energy values of the DNNP and the relaxed PCM scan are in better agreement with MTS-NNP. However, the smooth energy profile of the latter contrasts with the wiggly behavior of the DNNP profile, which is likely an outcome of the inconsistency discussed in the previous section. Finally, the relaxed scan in PCM reaches a plateau in the Te-\ce{Cl-} distance around 5 \AA~ indicative of an absence of interaction above this range. The absence of maximum is directly related to the missing explicit solvent molecules. 

Analysis of the N-Te-\ce{Cl-} angle (Fig. \ref{PMF}, B) further confirms the presence of directional chalcogen bonding in these PMF profiles in the short range regions. In longer distances, however, the directionality with \ce{Cl-} is lost with the angle dropping below 160$\degree$ similarly to MTS-NNP. In both the DNNP and AMOEBA simulations, the chalocogen bond becomes loose already around the Te-\ce{Cl-} distance of 3.5~\AA. The implicit solvent energy profile using the PCM predicts a shift of behavior at around 4.00~\AA.

Overall, the free energy profiles confirm the presence of a chalcogen bond between the benzotelluradiazole and \ce{Cl-}. The MTS-NNP result further highlights the strength and importance of this interaction especially in the short range regions (below 4.50~\AA). Yet, monitoring the PMF and angle is not sufficient to assess what is happening once the chalcogen bond is broken and to deeper understand the observed discrepancies between the MTS-NNP and other methods. The loss of directionality does not necessarily imply that \ce{Cl-} stop interacting with the \ce{Te} atom and with benzotelluradiazole at larger distances. While not measurable by UV-vis spectroscopy, other interactions might influence the chalcogen bond and affect its properties, namely anion-$\pi$ interactions and competition between \ce{Te}--\ce{Cl-} and \ce{Te}--\ce{O} chalcogen bonds with the anion and solvent molecules, respectively. The existence of competing interactions between benzotelluradiazole, THF and the chloride anion is further analyzed by the density distributions of oxygen atoms from solvent molecules and \ce{Cl-} around the heteroatom discussed in the next section.

\paragraph*{Competing non-covalent interactions in solution}
\begin{figure*}[!t]
    \centering
    \includegraphics[scale=0.62]{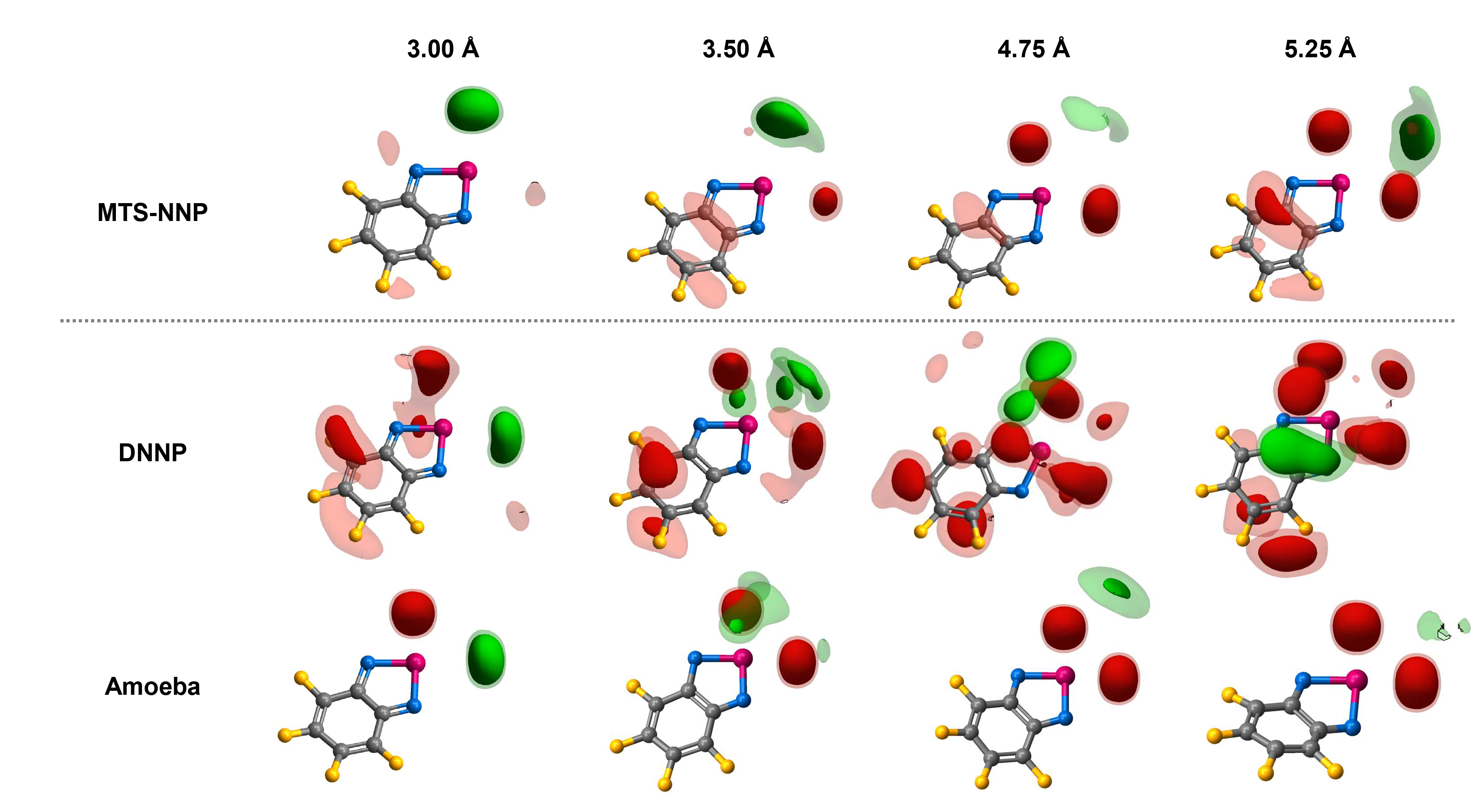}
    \caption{3D Density distribution of oxygen (red contour) and \ce{Cl-} (green contour) atoms around the  benzotelluradiazole. The density isocontours correspond to the density values of 0.005 (transparent) and 0.01 (opaque). A reference frame for density computation is defined by the N--Te--N atoms. C, N, F and Te atoms are depicted in gray, blue, yellow, and magenta, respectively. }
    \label{density}
\end{figure*} 

Fig. \ref{density} summarizes the density distributions of O and \ce{Cl-} atoms in the widows showcasing the largest differences between the MTS-NNP and DNNP PMF curves as given in Fig. \ref{PMF} (\textit{i.e.}, at Te--\ce{Cl-} distances of 3.00 \AA, 3.50 \AA, 4.75 \AA~and 5.25 \AA).

The MTS-NNP results confirms a directional chalcogen bond interaction between Te and \ce{Cl-}, which is apparent especially in the 3.00 \AA~distance window, which pictures the maximum of \ce{Cl-} density located close to the position of $\sigma$-hole. In the longer Te-\ce{Cl-} distances, interaction towards \ce{Cl-} weakens and the affinity towards the oxygen atoms of THF is amplified instead. The windows at Te-\ce{Cl-} distance of 4.75 and 5.25 \AA~finally demonstrate complete replacement of Te and \ce{Cl-} chalcogen bond by directional chalcogen bonds towards oxygen atoms, which occupy both $\sigma$-hole regions. The directionality of the newly-formed Te-O bonds is further confirmed by the N-Te-O angle distributions (see Fig. S15 in the SM). This behavior further stresses the importance of Te-\ce{Cl-} chalcogen bonds in the short range with its gradual replacement by Te-\ce{O} chalcogen bonds at the Te-\ce{Cl-} distances above 4.00 \AA, which explains the observed trends in the N-Te-\ce{Cl-} angle distributions. The density distribution at the 5.25 \AA~window depicts the transition from the Te-\ce{Cl} complex to the fully solvated \ce{Cl-} corresponding to the maximum of the PMF profile located around 5.2~\AA~(see Fig. \ref{PMF}A).

At 3.00 \AA, the density distributions with DNNP and AMOEBA already differ with those obtained with the MTS-NNP. They promote an interaction with \ce{Cl-} but also one with a THF oxygen atom, which is not present using MTS-NNP. At Te-\ce{Cl-} distances larger than 3.00 \AA, the directional interaction with \ce{Cl-} is entirely replaced by the Te--O chalcogen bond interactions. With AMOEBA, the interaction between \ce{Te} and \ce{Cl-} decreases progressively to ultimately vanish at distances greater than 5.25 \AA. This behavior, which is aligned with the PMF, demonstrates the limitations of AMOEBA for describing chalcogen bonds involving tellurium. The density distributions with DNNP are rather nonphysical exhibiting a too strong affinity for the solvent and to some extent too strong anion-$\pi$ interactions that are not observed in the MTS-NNP simulations. These artifacts arise from a poor extrapolation of the neural network potential in these regions and further emphasize the importance of its careful validation.

In general, the density distributions presented here show that the two $\sigma$-hole regions on the tellurium atom are good donors of chalcogen bond towards both the \ce{Cl-} anion and the THF oxygen atom. Benzotelluradiazole therefore interacts not only with the \ce{Cl-} but also with the solvent, which is not apparent from implicit solvent computations. The presence of the explicit solvent molecules is thus essential to properly describe the breaking of the chalcogen bond arising from competing interactions. More importantly, the solvent containing atoms in possession of lone pairs (\textit{e.g.}, O, N) can directly compete with the chalcogen bond acceptors and influence the interaction strength. The impact of the different solvents on the chalcogen bonded molecules was already observed in the formation of aggregates and crystallization.\cite{BIOT2020213243} For instance, similar tellurium-containing compound crystallizes in a form of Te-\ce{O-} bonded macrocycles with a stoichiometry dictated directly by the solvent effects.\cite{Ho2016} Interestingly, THF molecules were even found inside the formed cavities further advocating for their interaction with tellurium. Chalcogen bonds between selenium containing species and oxygen and nitrogen atoms of solvent molecules were also confirmed by NMR measurements. \cite{D0NJ04647G}   

Overall, this section provides key chemical and technical insights regarding both the nature and importance of chalcogen bonds and the reliability of their condensed phase simulations using distinct potentials, ranging from polarizable force fields to neural network surrogates for hybrid DFT. This comparison demonstrates the larger reliability and robustness of the MTS-NNP approach and the difficulties of the DNNP and polarizable force field in the description of challenging charge transfer complexes. From the chemical perspective, it is evident that chalcogen bonds will unavoidably form with the solvent molecules and will consequently enforce a more structured solvation shell near the solute which cannot be captured by implicit solvent models.  

\section{Conclusions}
This computational work investigates the persistence of chalcogen bonds between benzotelluradiazole  and the \ce{Cl^-} anion solvated in tetrahydrofuran. To ensure  both quantum chemical accuracy and statistical convergence, we simulate the complex multielement systems made of 599 atoms relying upon Behler-Parrinello neural network potentials.   
We compare the performance of a MTS-NNP and a direct NNP which are both trained to reproduce hybrid DFT energies and forces. The two variants show an impressive stability along the nanosecond timescale molecular dynamic simulations. The stability of the baseline-free DNNP is especially encouraging, however, still requires significant effort to improve its generalization properties. From the technical perspective, we demonstrate the versatility of the MTS-NNP approach combining baselined and direct NNP in the simulation of system incorporating several types of competing non-covalent interactions.

The application of this method brings important chemical insight into behavior of chalcogen bonded molecules in the explicit solvent, which would not be possible to obtain with standard state-of-the-art approaches, \textit{i.e.}, \textit{ab initio} molecular dynamics or polarizable force fields. Specifically, we confirm the persistence of the chalcogen bond in solution and observe an important competition between the chalcogen bond towards \ce{Cl-} and the solvent oxygen atoms. While the Te-\ce{Cl-} chalcogen bond is a dominant non-covalent interaction in the short range region up to Te-\ce{Cl-} distances of 4.00 \AA, the interaction between Te atom and oxygen of THF is prevailing in larger distances. This behavior coincides with the loss of directionality of the \ce{Te}--\ce{Cl-} interaction and highlights the non-negligible impact of oxygen or nitrogen containing solvents on the topology of the chalcogen bond. 

This work demonstrates that neural networks can be successfully applied to model the energy profiles of complex molecular systems interacting with the solvent. They provide important atomistic resolution of the solvation mechanism which helps rationalizing and harvesting the nature and behavior of chalcogen bonds in solution. Similar approaches would be useful for assessing the role of Lewis basic solvents in the co-crystallization of chalcogen-bearing compounds and the potential reduction of catalysis activity or association constants induced by the solvent. More generally, it does open efficient and reliable routes to simulate homogeneous organocatalysts, host-guest systems and other molecular systems exploiting a subtle balance of non-covalent interactions.  

\section*{Supplementary material}

See the supplementary material for computational details, information on AMOEBA parametrization, NNP validation, convergence of PMF and supplementary plots.

\begin{acknowledgments}
The authors are grateful to the EPFL for financial support and computational resources. V. J. acknowledges the funding from the Swiss National Science Foundation (SNSF, No. 514149), and F. C. and C. C. from the European Research Council (ERC, Grant Agreement No. 817977) within the framework of European Union's H2020. The National Center  of  Competence  in  Research  (NCCR) “Sustainable chemical process through catalysis (Catalysis)” of SNSF is acknowledged for financial support of R. L.  The authors thank Raimon Fabregat for the assistance with graphical material, Daniel Hollas for the help with Abin set up and Kevin Rossi for scientific discussions. 
\end{acknowledgments}

\section*{Data Availability Statement}

The xTB-ase calculator is available on Github at \url{https://github.com/lcmd-epfl/xtb_ase_io_calculator}. The data and the neural networks parameters will be freely available on the Materials Cloud.

\section*{References}

\bibliography{bib}% Produces the bibliography via BibTeX.

\end{document}

\title{{\sc Supplementary Material}\texorpdfstring{\\}{}
Assessing the persistence of chalcogen bonds in solution\texorpdfstring{\\}{}
with neural network potentials}

\author{V. Jur\'{a}skov\'{a}}
\affiliation{Laboratory for Computational Molecular Design, Institute of Chemical Sciences and Engineering,
\'{E}cole Polytechnique F\'{e}d\'{e}rale de Lausanne, 1015 Lausanne, Switzerland}
\author{F. C\'{e}lerse}
\affiliation{Laboratory for Computational Molecular Design, Institute of Chemical Sciences and Engineering,
\'{E}cole Polytechnique F\'{e}d\'{e}rale de Lausanne, 1015 Lausanne, Switzerland}
\author{R. Laplaza}
\affiliation{Laboratory for Computational Molecular Design, Institute of Chemical Sciences and Engineering,
\'{E}cole Polytechnique F\'{e}d\'{e}rale de Lausanne, 1015 Lausanne, Switzerland}
\affiliation{National Center for Competence in Research-Catalysis (NCCR-Catalysis), \'Ecole Polytechnique F\'ed\'erale de Lausanne, 1015 Lausanne, Switzerland.}
\author{Clemence Corminboeuf}
\email{clemence.corminboeuf@epfl.ch}
\affiliation{Laboratory for Computational Molecular Design, Institute of Chemical Sciences and Engineering,
\'{E}cole Polytechnique F\'{e}d\'{e}rale de Lausanne, 1015 Lausanne, Switzerland}
\affiliation{National Centre for Computational Design and Discovery of Novel Materials (MARVEL),
\'{E}cole Polytechnique F\'{e}d\'{e}rale de Lausanne, 1015 Lausanne, Switzerland}
\affiliation{National Center for Competence in Research-Catalysis (NCCR-Catalysis), \'Ecole Polytechnique F\'ed\'erale de Lausanne, 1015 Lausanne, Switzerland.}

\date{\today}

\maketitle
\onecolumngrid
\tableofcontents

\newpage
\section{Computational details}
\subsection{Electronic structure description}
GFN0--xTB \cite{pracht2019robust} with periodic boundary conditions as implemented in xTB 6.2 \cite{https://doi.org/10.1002/wcms.1493} is used both for preliminary sampling and as a baseline potential for NNP simulations. GFN0--xTB is a non--iterative member of the extended tight-binding family of Hamiltonians developed by Grimme et al., which makes the approach significantly faster compared to self--consistent semiempirical methods. Reference energies and forces are computed at the PBE0--D3BJ\cite{ernzerhof1999assessment,grimme2011effect} level of theory as implemented in CP2K 6.1.\cite{kuhne2020cp2k} We use the TZV2P--MOLOPT basis set for \ce{H}, \ce{C}, \ce{N}, \ce{O}, \ce{F} and \ce{Cl},\cite{vandevondele2007gaussian} and a DZVP--MOLOPT--SR for \ce{Te} with a Goedecker-Teter-Hutter pseudopotential (GTH PBE)\cite{krack2005pseudopotentials} for all elements. The plane--wave cutoff is set to 700 Ry with a relative cutoff of 70 Ry. All computations employ a Coulomb operator truncated at R = 9 \AA~ and the auxiliary density matrix method with a cpFIT3 fitting basis set for \ce{H}, \ce{C}, \ce{N}, \ce{O}, \ce{F} and \ce{Cl} and  FIT9 for \ce{Te}.\cite{guidon2010auxiliary} PBE--D3BJ wave functions were used as the initial guess for the PBE0--D3BJ computations to speed up the convergence.\cite{adamo1999toward,grimme2011effect} PBC are taken into account with a cubic box of 18.428 \AA~ and a total charge of -1.0\textit{e} for the simulation cell. The free energy curves in implicit solvent are computed using the PCM,\cite{Tomasi2005} SMD,\cite{doi:10.1021/jp810292n} and COSMO-RS\cite{doi:10.1021/j100007a062} models. Structures for the computations are generated by a relaxed scan along the Te-\ce{Cl-} distance in a range from 2.0 to 7.75 \AA~with a step of 0.25 \AA~at the PBE0-D3BJ/aug-cc-pVTZ level with the solvent approximated by PCM and SMD, respectively, as implemented in Gaussian 16.\cite{g16} Free energy curves for the PCM and SMD models are constructed from 26 structures computed at PBE0-D3BJ/aug-cc-pVTZ single points with the respective solvent model using the quasi-harmonic correction to entropy\cite{Grimme2012} and enthalpy\cite{Li2015} as implemented in the GoodVibes program.\cite{good_vibes} Energy estimates and COSMO-RS solvation energies are computed at the PBE0-dDsC/TZ2P level, as implemented in the ADF package ,\cite{TeVelde2001} using the same geometries as in the PCM computations. Free energy corrections are obtained using the PBE0-D3BJ/aug-cc-pVTZ level along with the quasi-harmonic correction to entropy and enthalpy as in previous computations. 

\subsection{Classical force fields simulations}

To compare the NNPs performance with state-of-the-art force fields, we perform the umbrella sampling simulations with polarizable (AMOEBA) force field. Since the suitable AMOEBA parameters for THF and benzotelluradiazole do not exist, both systems were parametrized before proceeding with the simulations using the following protocol and mixture: one benzotelluradiazole molecule, 45 THF molecules, one \ce{Cl-} anion and one \ce{Na^+} cation. \ce{Na+}, which can influence the PMF, is constrained at a distance of 10 \AA~with a force constant of 100 kcal/mol~$cdot$\AA$^2$ applied in all simulations. The molecular dynamics simulations are carried out using the Tinker 8.8 software.\cite{rackers2018tinker} The equations of motion are integrated by the RESPA multiple--time--step integrator with a time step of 2 fs. The Smooth Particle Mesh Ewald algorithm is employed with a grid dimension of 24 \AA~ $\times$ 24 \AA~ $\times$ 24 \AA~ to calculate forces in the reciprocal space. The Ewald cutoff is set to 7 \AA~ and the Van der Waals cutoff to 9 \AA. The NVT thermodynamic ensemble is used with the temperature maintained to 298.15 K by the Bussi thermostat. For the AMOEBA force field, the convergence criteria for multipoles are set to 10$^{-5}$. We define the same windows and spring constants as for the NNPs simulations, but the simulation time per window is increased to 10 ns. The first two nanoseconds are kept for equilibration and not considered in the analysis. Finally, the PMF is reconstructed and the statistical error estimated in both cases using the WHAM implementation of Alan Grossfield.\cite{grossfield2003wham} 

\newpage
\section{AMOEBA parametrization}
AMOEBA parameters are generated for the benzotelluradiazole and THF molecules according to the procedure described by Ren \textit{et al.}\cite{wu2012automation} with the following main steps:
\begin{enumerate}
    \item The geometry of an initial structures are optimized at the MP2/6-311G** level using the Gaussian 09 software.
    \item The distributed Multipole Analysis is performed to generate an initial set of atomic multipoles.
    \item The multipoles are refined by fitting to the molecular electrostatic potential obtained from a single point calculation at the MP2/aug--cc--pvtz level of theory.
    \item Van der Waals, bond, angle, dihedral, and out of plane bending parameters are assigned using the Tinker \textit{valence} program. 
    \item Structures containing one benzotelluradiazole with 45 THF molecules are built using the Tinker \textit{xyzedit} program and minimized with a RMS of 1.0 kcal/mol using the Tinker \textit{minimize} program. 15 ps NVT following by 2 ns NPT simulations  are performed to check the consistency of the simulation parameters.
\end{enumerate}

\clearpage
\includepdf[pages={1,{},2-5}]{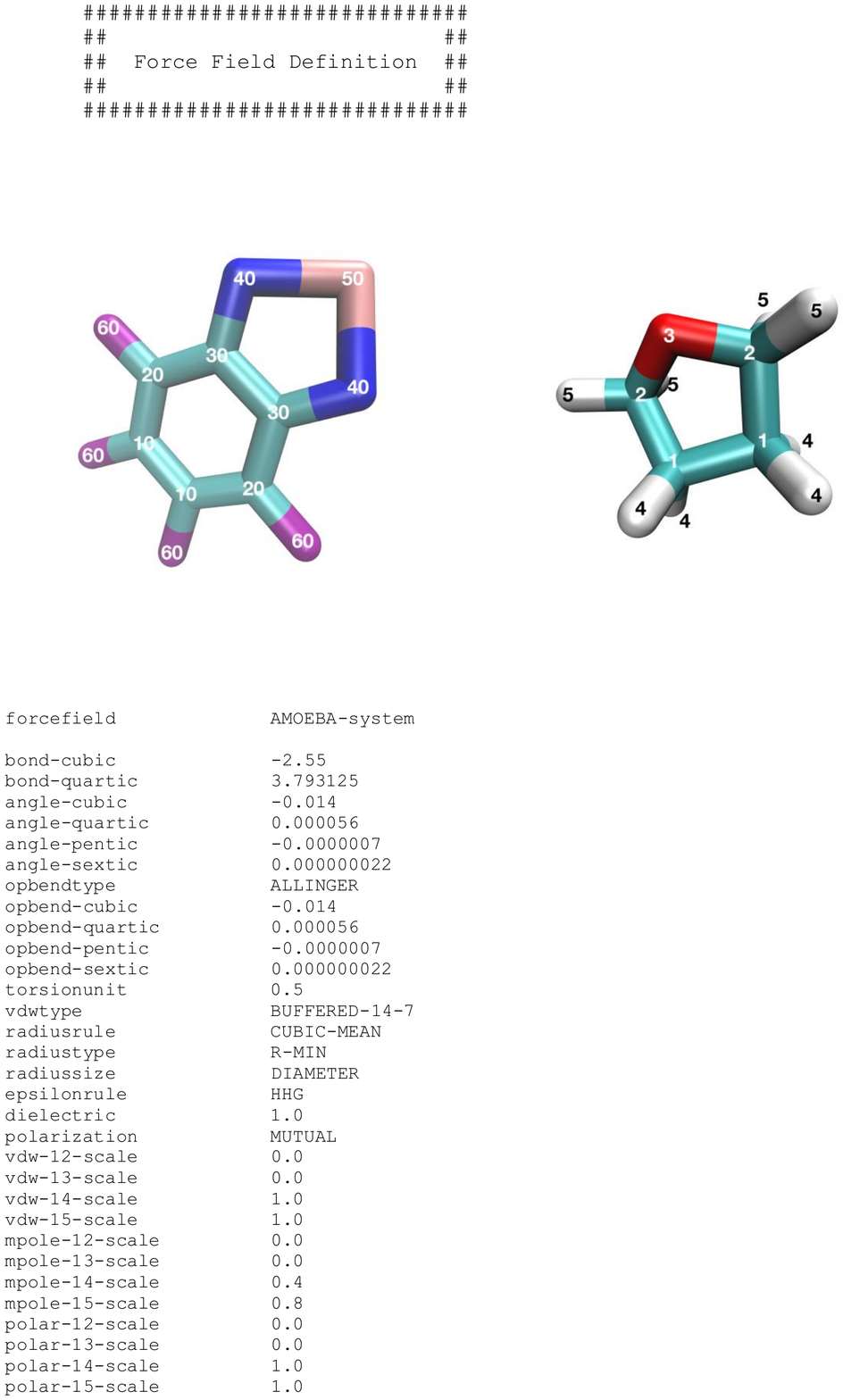}

%\newpage
\subsection{Parameter validation}
\begin{figure*}[!htp]
    \centering
    \includegraphics{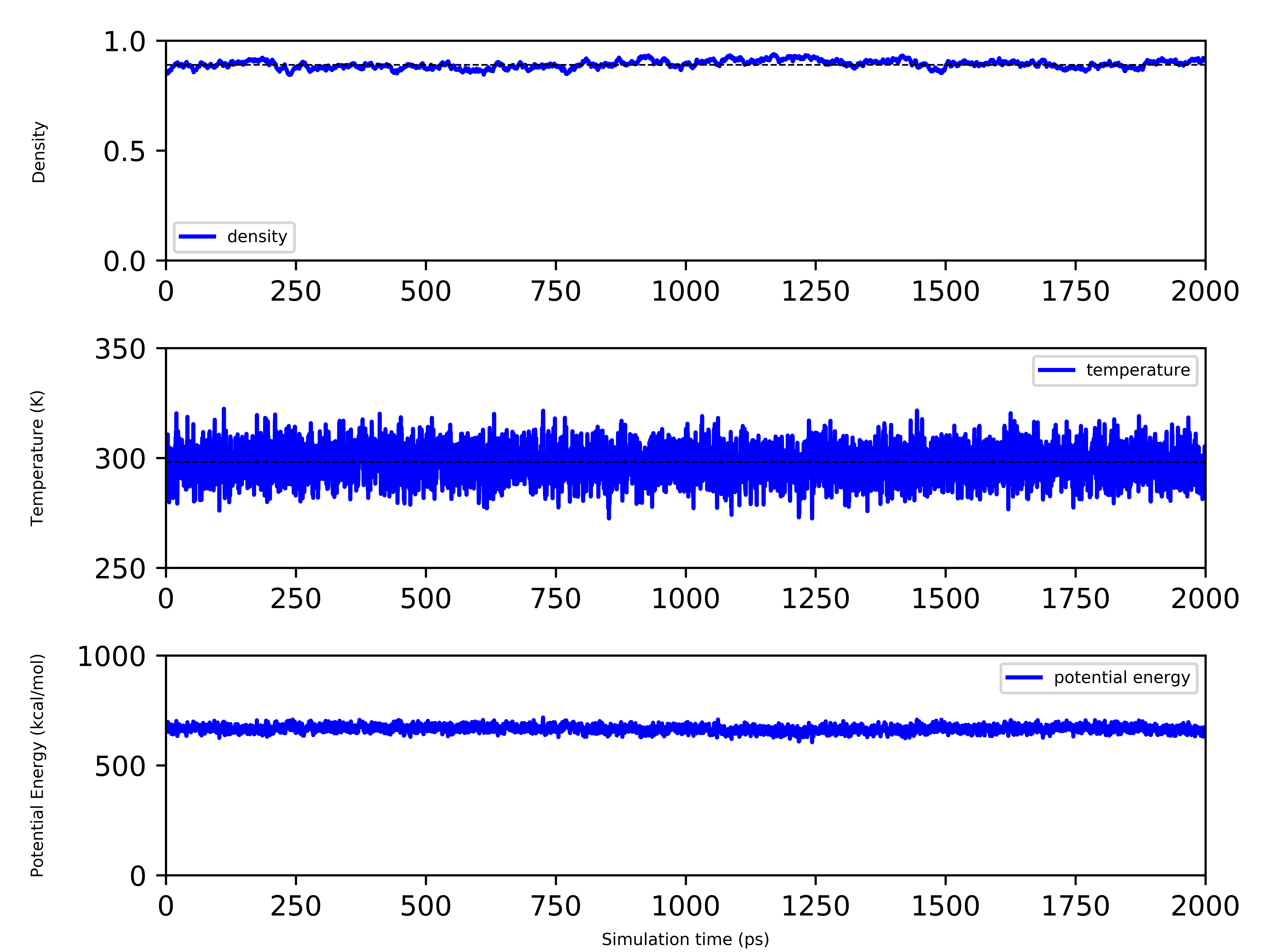}
    \caption{Evaluation of the density, temperature and potential energy as a function of the simulation time. The black dashed line depicts the targeted values for density (exp. 0.89 g/cm$^3$) and temperature (298.15 K).}
    \label{proof_par_amoeba}
\end{figure*}

\newpage
\section{Symmetry functions selection}

\begin{figure*}[!htp]
    \centering
    \includegraphics{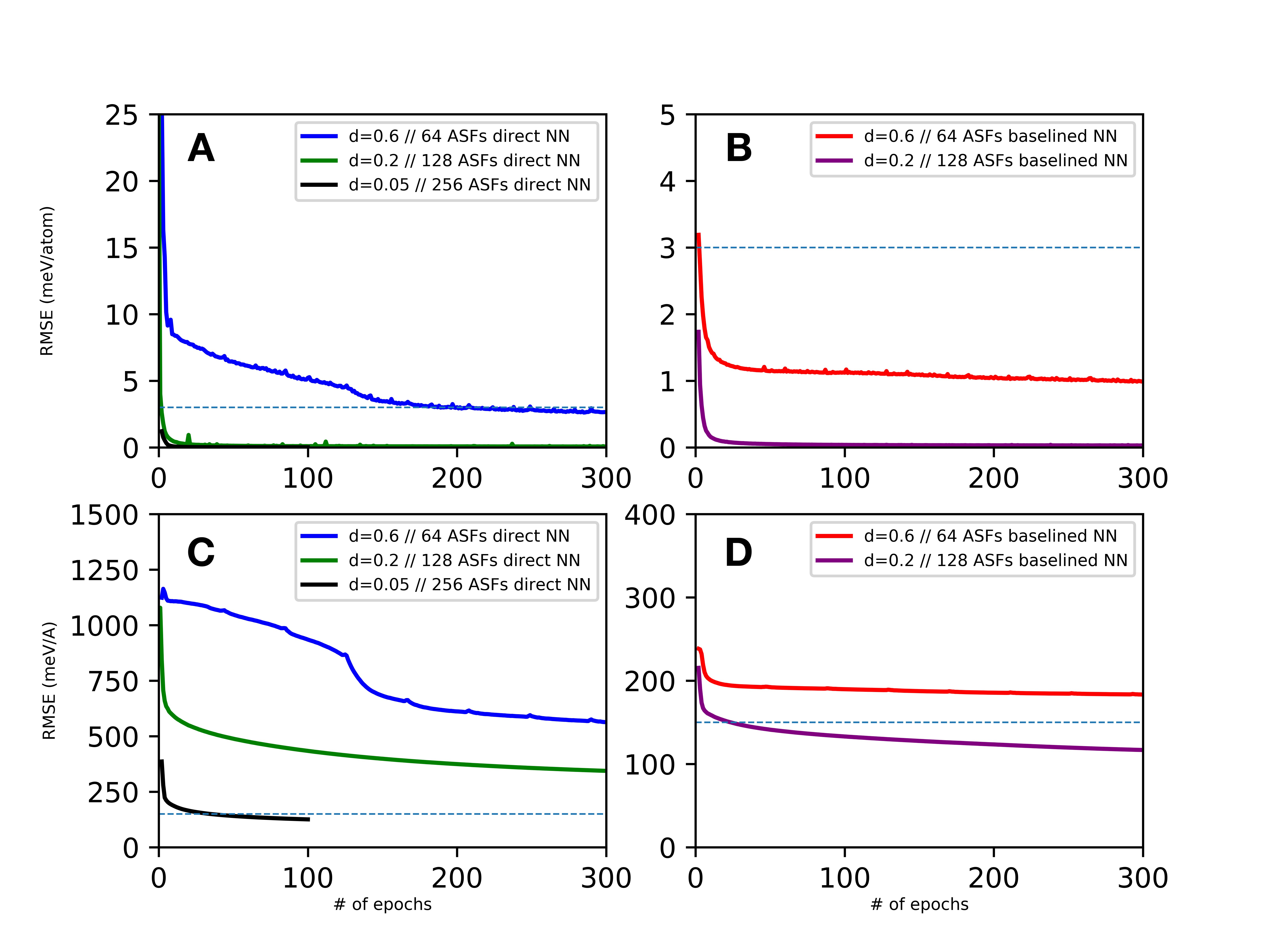}
    \caption{Training RMSEs as a function of training epoch for energy (A and B) and forces (C and D) for different numbers of ACSFs. The dashed line represents the accuracy threshold, which is set to 3 meV/atom and 150 meV/\AA~ for energy and forces, respectively.}
    \label{NNP-train}
\end{figure*}

\newpage
\section{NNP validation}

\subsection{Training errors of the test set}

\begin{figure*}[!htp]
    \centering
    \includegraphics[scale=0.65]{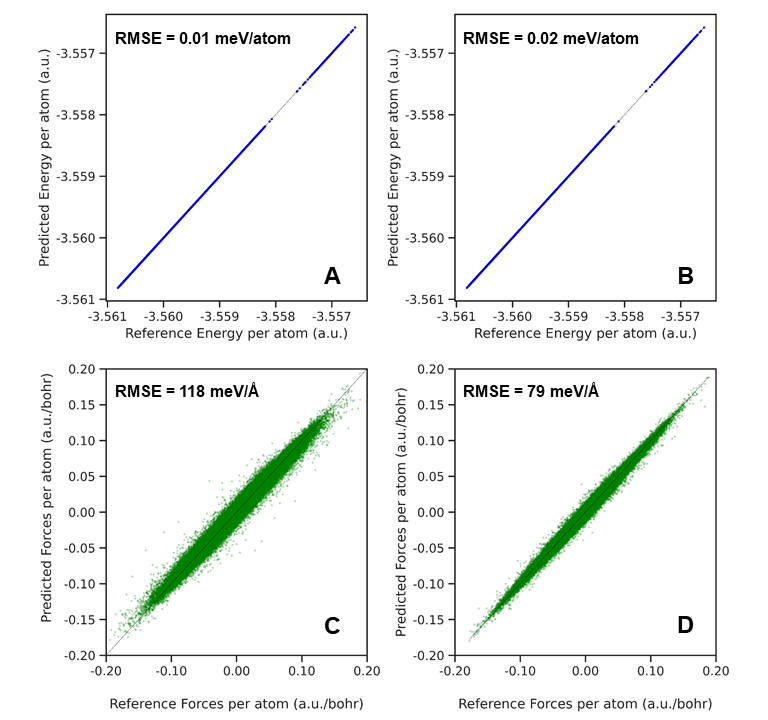}
    \caption{Parity plots for direct and baselined NNPs for energy (direct in A and baselined in B) and forces (direct in C and baselined in D). Results are reported after 300 training epochs for the baselined and 100 for the direct NN. Data are obtained for the 80\% of structures contained in the training set.}
    \label{RMSE-testset}
\end{figure*}

\newpage
\subsection{Forces parity plots per element}

\begin{figure*}[!htp]
    \centering
    \includegraphics[scale=0.37]{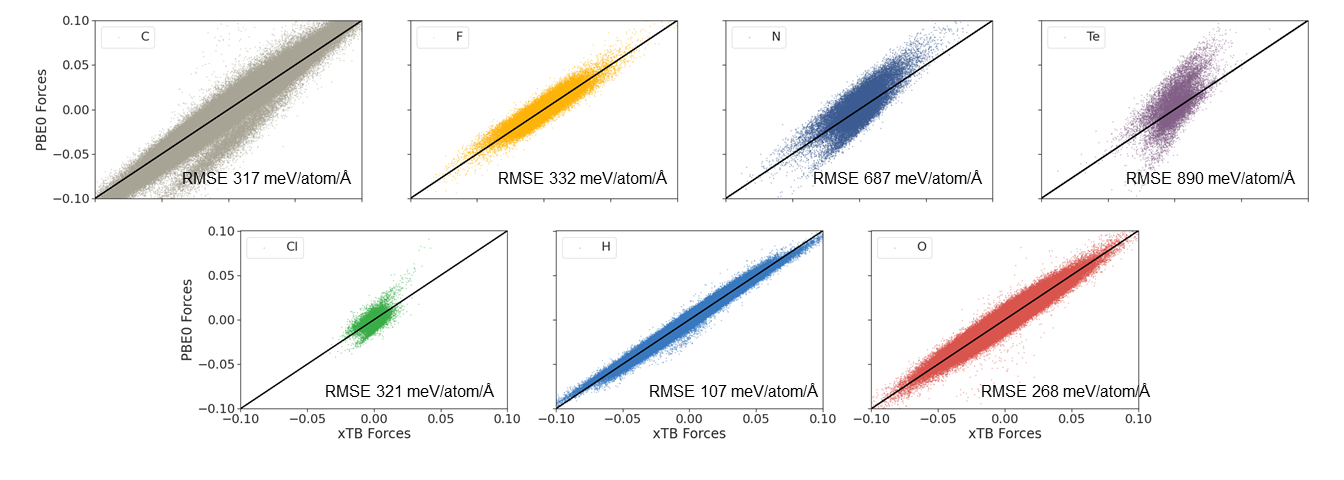}
    \caption{Force parity plots per element, xTB vs. PBE0-D3BJ. Data from the test set.}
    \label{RMSE_xtb}
\end{figure*}
\begin{figure*}[!htp]
    \centering
    \includegraphics[scale=0.37]{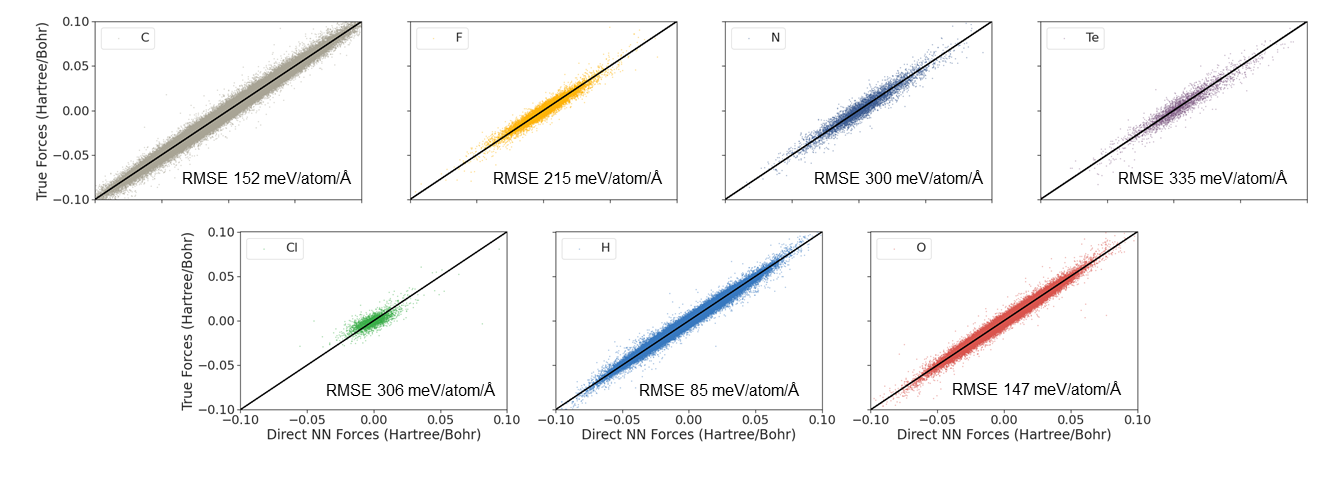}
    \caption{Force parity plots per element, direct NN vs. PBE0-D3BJ. Data from the test set.}
    \label{RMSE_direct}
\end{figure*}
\begin{figure*}[!htp]
    \centering
    \includegraphics[scale=0.37]{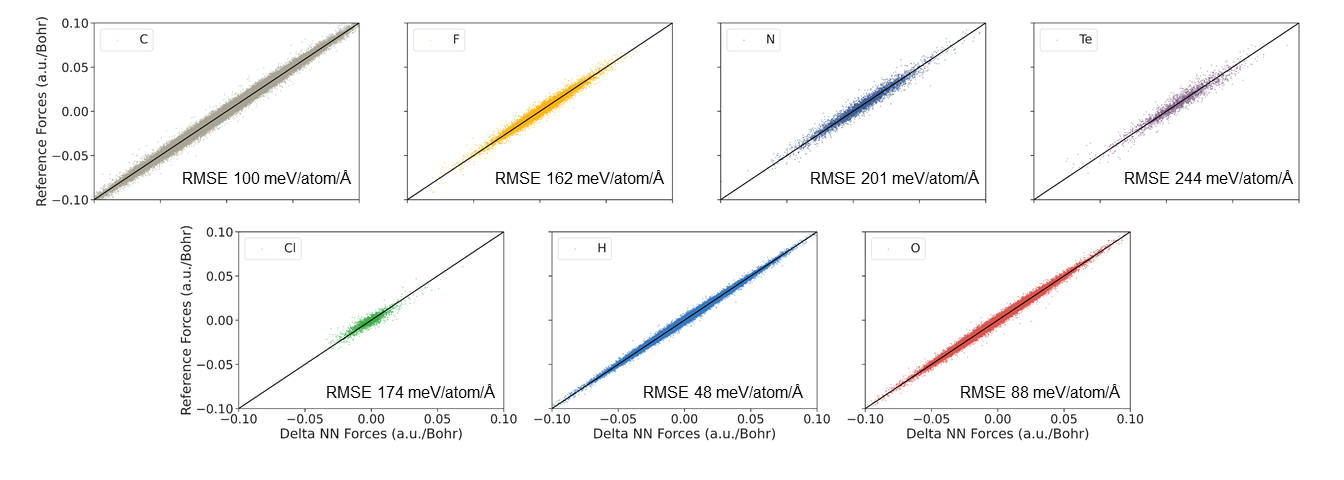}
    \caption{Force parity plots per element, baselined NN vs. PBE0-D3BJ. Data from the test set.}
    \label{RMSE_delta}
\end{figure*}

\newpage
\section{Potential of Mean Force: Convergence verification}

\subsection{AMOEBA}
\begin{figure*}[!htp]
    \centering
    \includegraphics[scale=1]{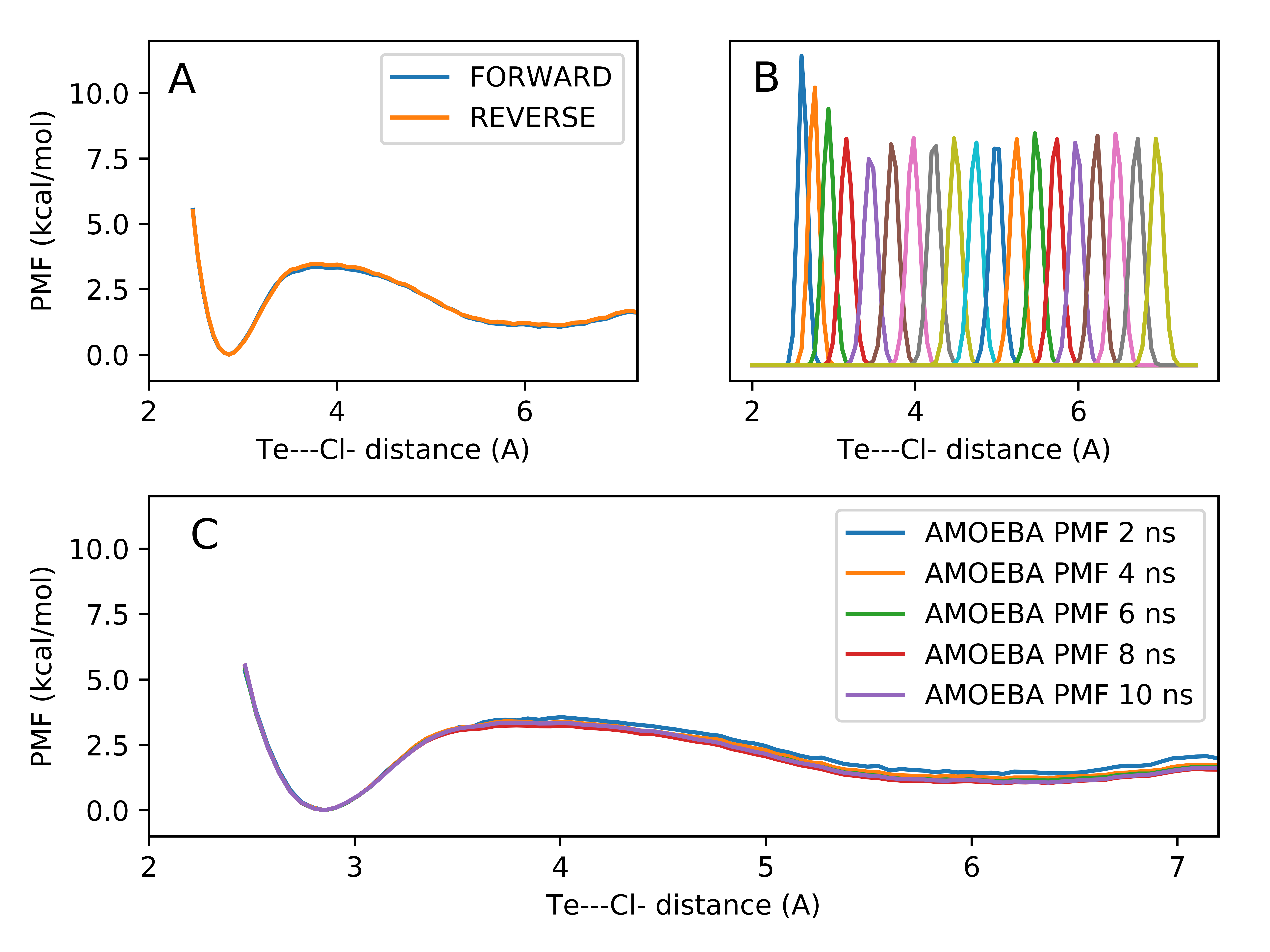}
    \caption{AMOEBA simulation: A: Forward/reverse PMFs; B: Histograms of the collective variable depicted for each window ranging from 2.5 to 7 \AA~ with a width of 0.25 \AA; C: Convergence of PMF as a function of the simulation time allowed per window.}
    \label{PMF-conv-A18}
\end{figure*}
\newpage
\subsection{DNNP}
\begin{figure*}[!htp]
    \centering
    \includegraphics[scale=1]{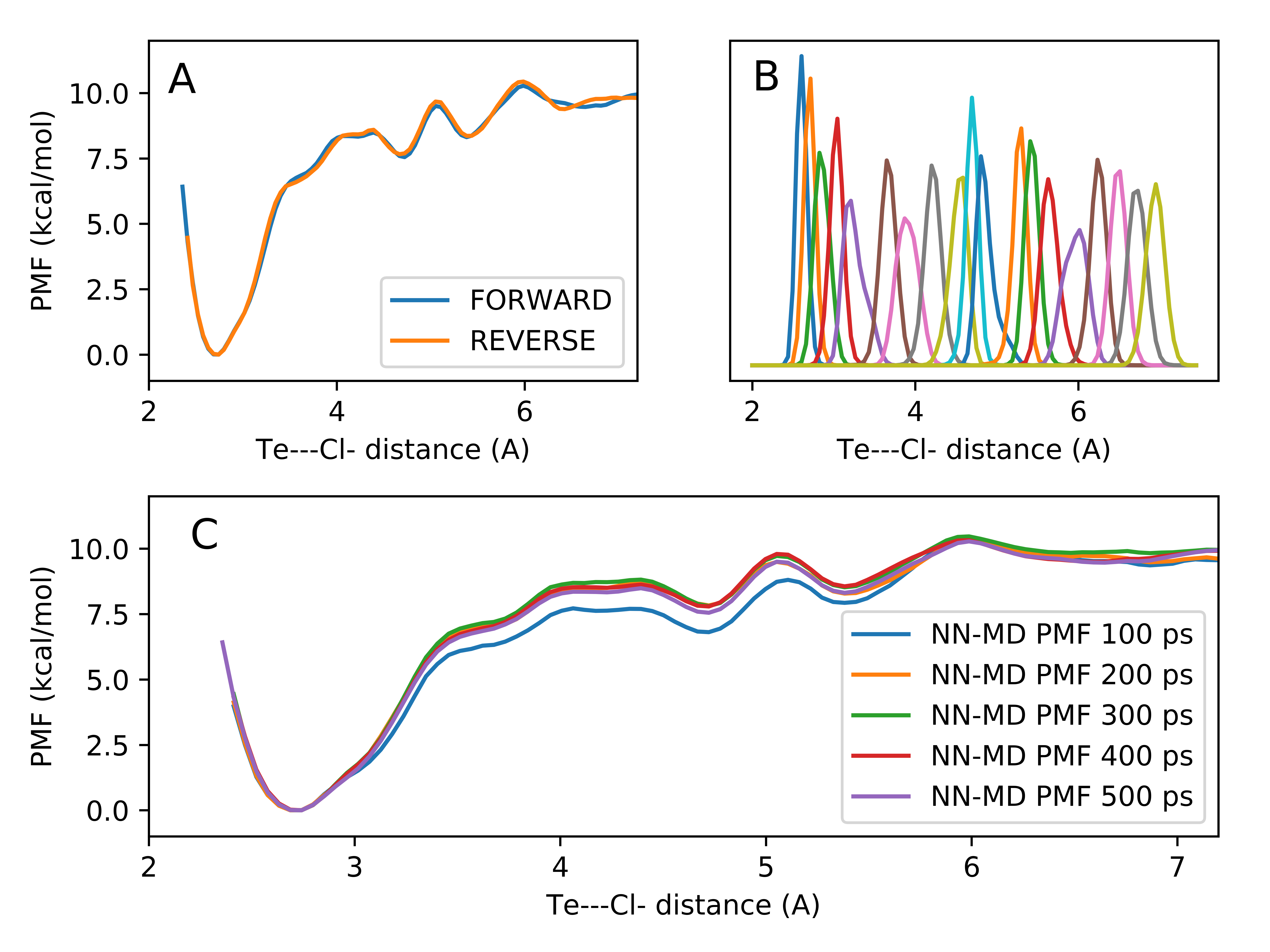}
    \caption{DNNP simulation: A: Forward/reverse PMFs; B: Histograms of the collective variables depicted for each window ranging from 2.5 to 7 \AA~ with a width of 0.25 \AA; C: Convergence of PMF as a function of the simulation time per window.}
    \label{PMF-conv-dir}
\end{figure*}
\begin{figure*}[!htp]
    \centering
    \includegraphics[scale=1]{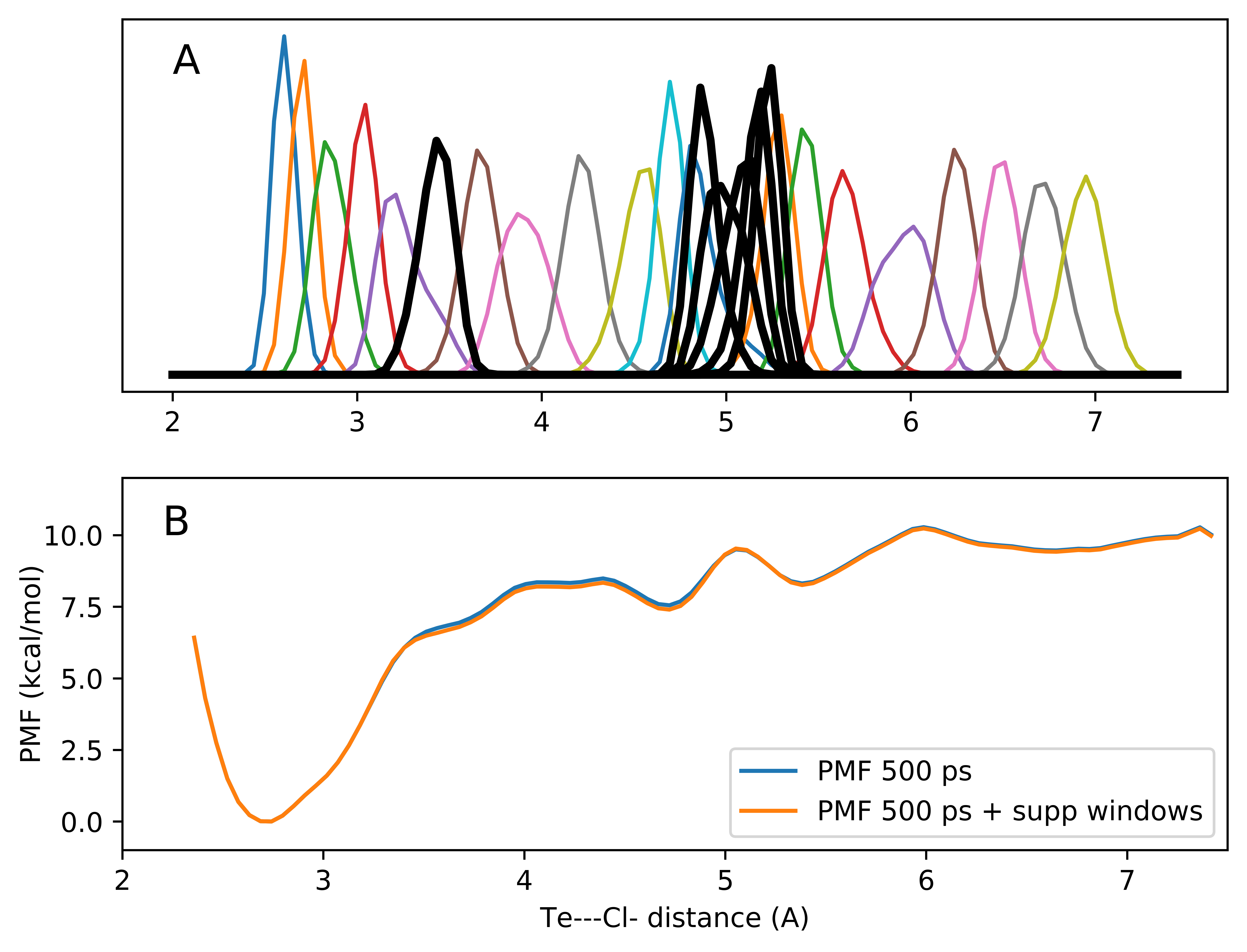}
    \caption{DNNP simulation: A: Histograms from figure~\ref{PMF-conv-dir} subgraph B with supplementary umbrellas (colored in black, 3.50, 5.00, 5.05, 5.10, 5.15 and 5.20 \AA); B: Comparison of the new PMF and the old PMF from Figure~\ref{PMF-conv-dir} subgraph A.}
    \label{PMF-conv-A18_2}
\end{figure*}
\clearpage

\newpage
\subsection{MTS-NNP}
\begin{figure*}[!htp]
    \centering
    \includegraphics[scale=1]{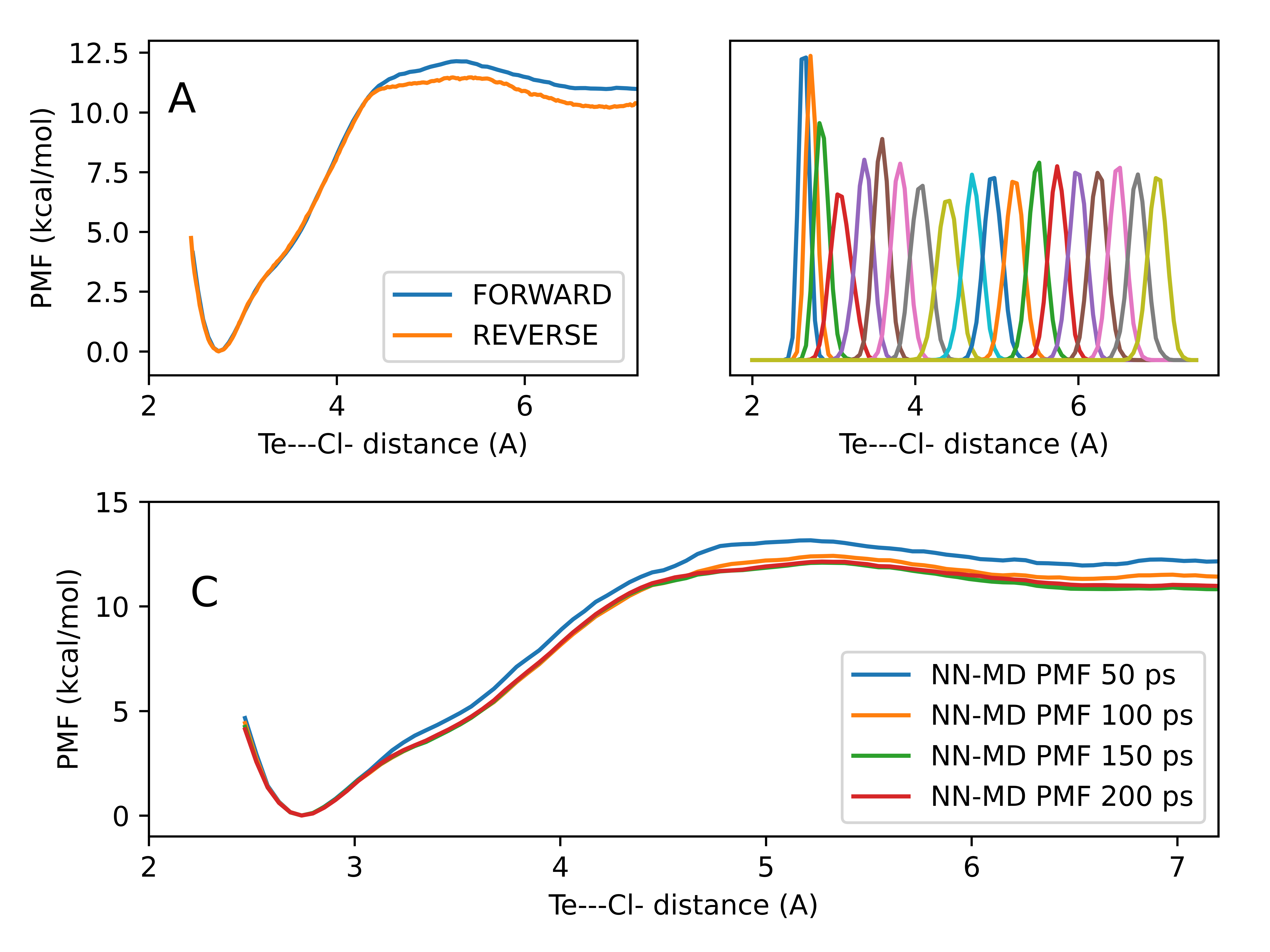}
    \caption{MTS-NNP simulation: A: Forward/reverse PMFs; B: Histograms of the collective variable depicted for each window ranging from 2.5 to 7 \AA~ with a width of 0.25 \AA; C: Convergence of PMF as a function of the simulation time per window.}
    \label{PMF-conv-mts}
\end{figure*}

\newpage
\section{Supplementary/Supplementary plots}
\subsection{Comparison of implicit solvent models}

\begin{figure*}[!htp]
    \centering
    \includegraphics[scale=0.9]{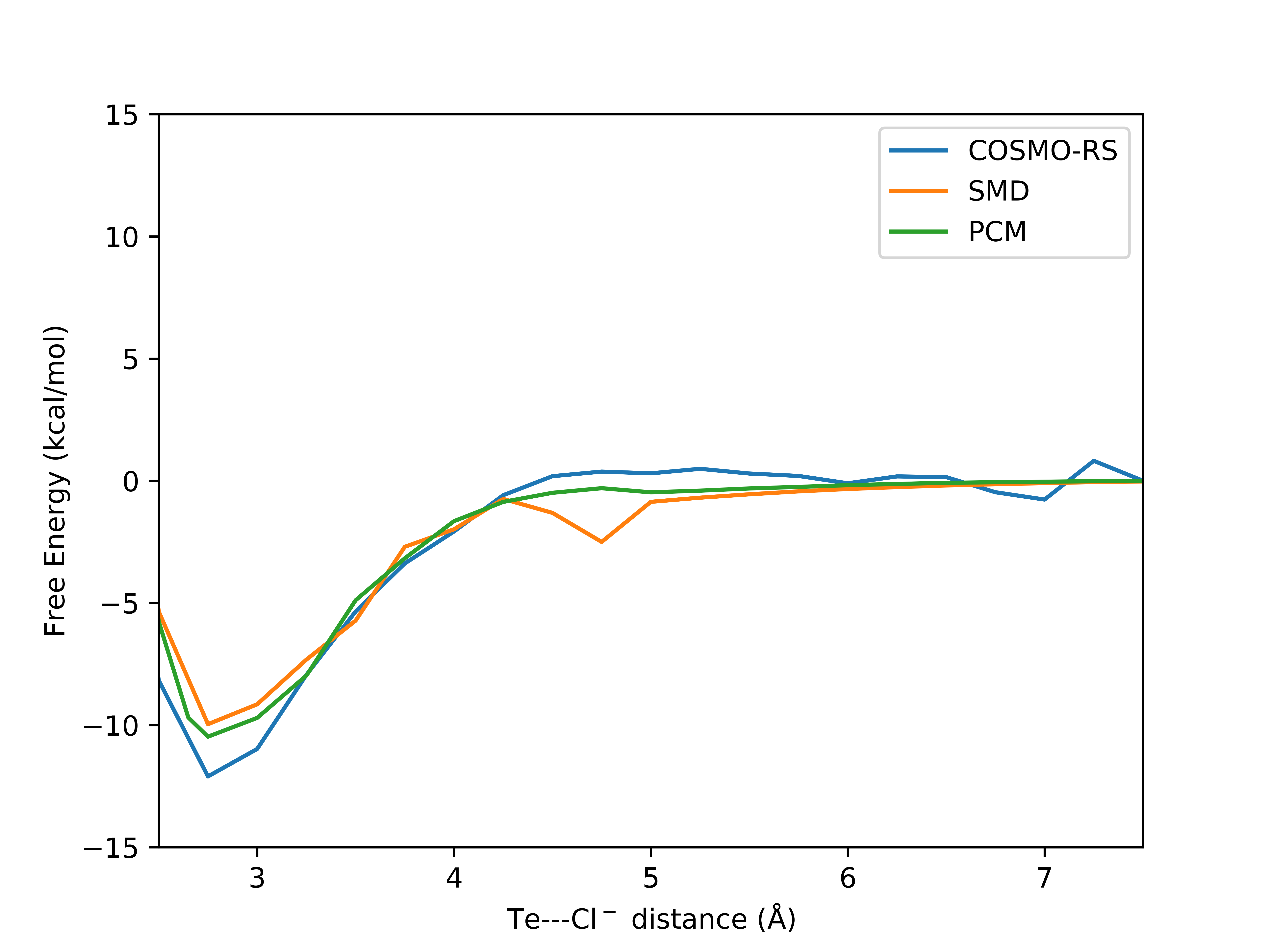}
    \caption{Free energy profile computed for structures from a relaxed scan along the Te--\ce{Cl-} distance using PCM, SMD and COSMO-RS implicit solvent models.}
    \label{implicit_comparison}
\end{figure*}

\newpage
\subsection{Extrapolation warnings}
\begin{figure*}[!htp]
    \centering
    \includegraphics[scale=0.8]{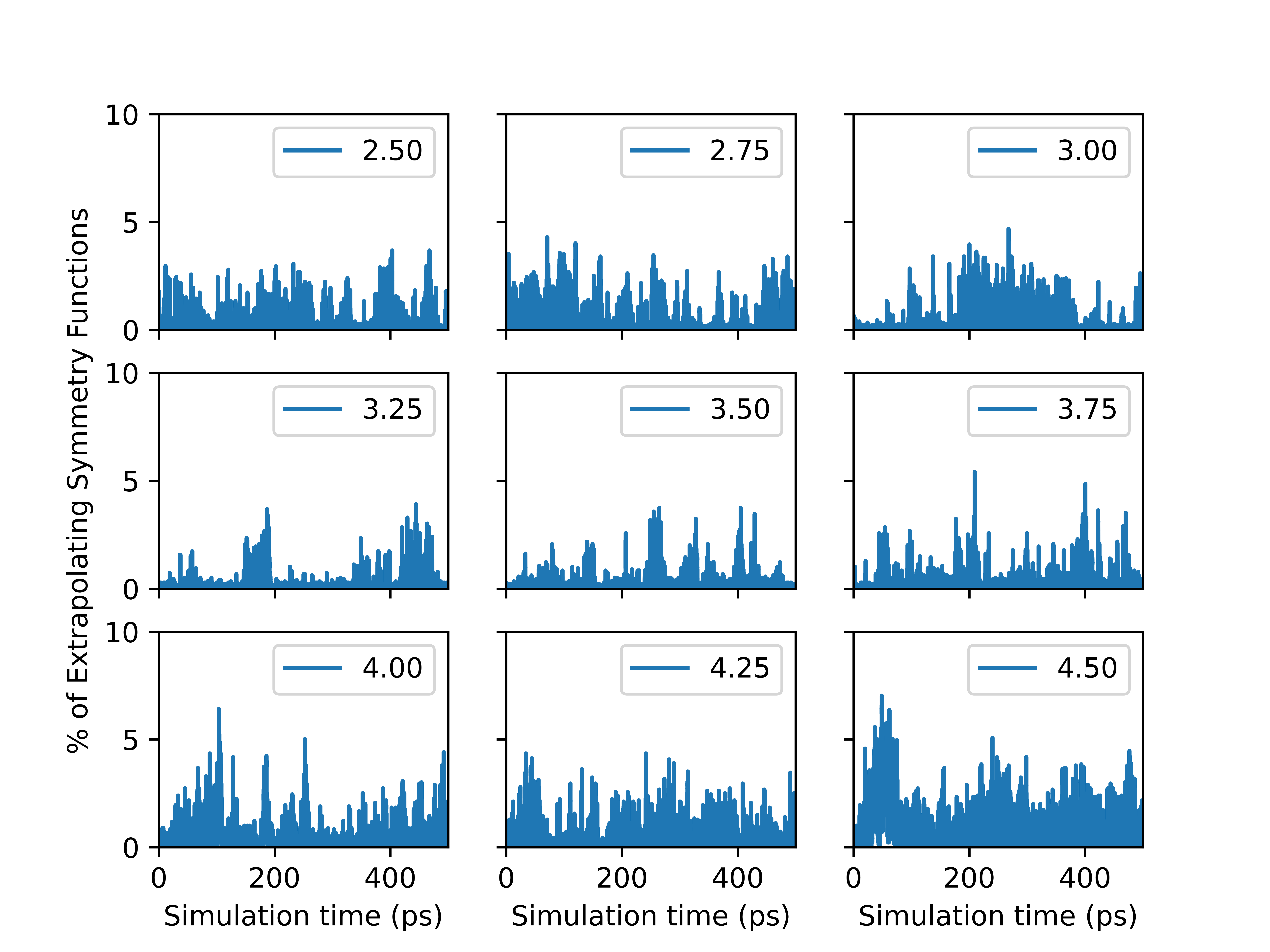}
    \includegraphics[scale=0.8]{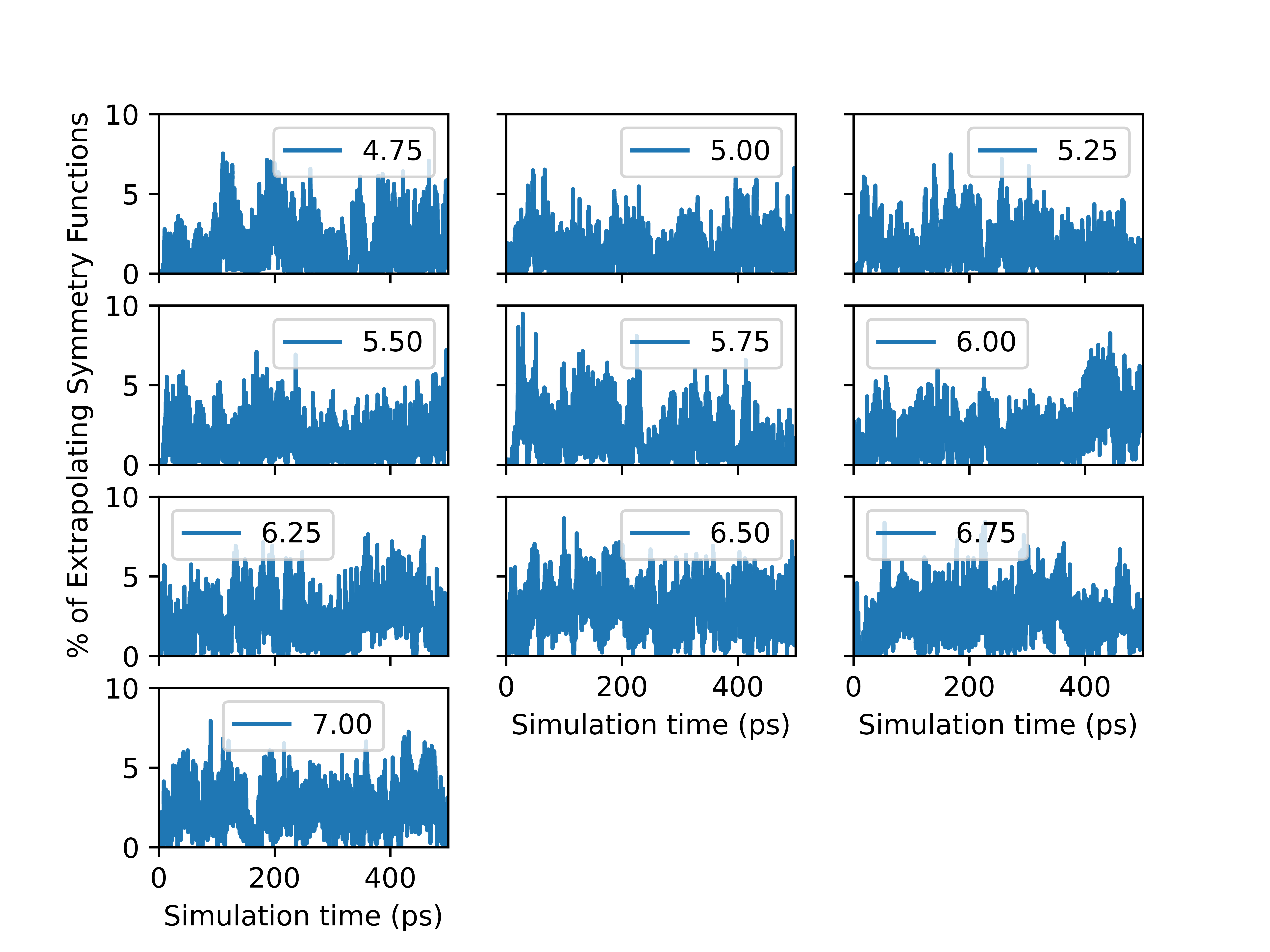}
    \caption{Extrapolation warnings for each window (label corresponds to the restrained distance) obtained with the DNNP US simulation.}
    \label{ew-direct-direct}
\end{figure*}
\begin{figure*}[!htp]
    \centering
    \includegraphics[scale=0.8]{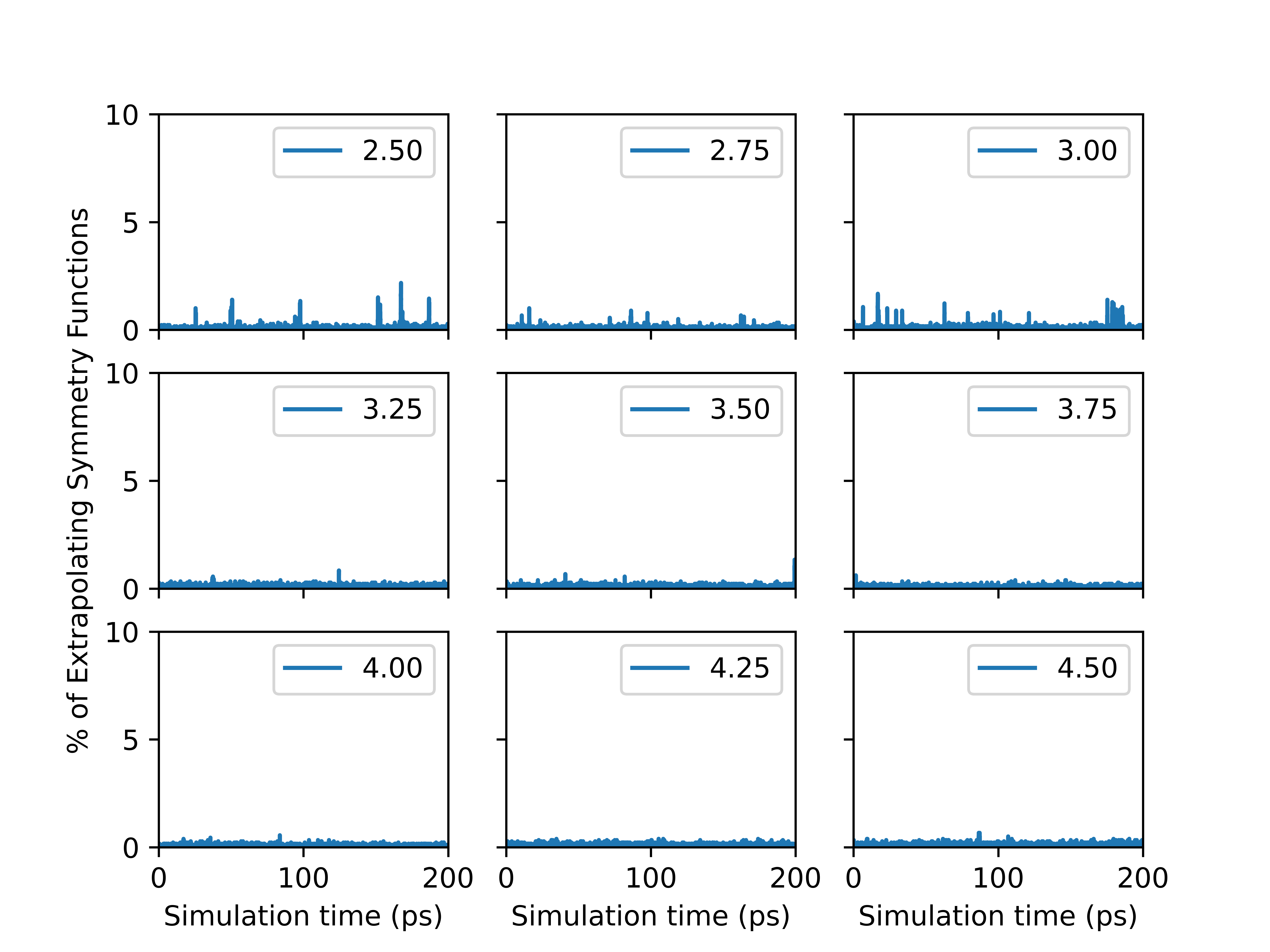}
    \includegraphics[scale=0.8]{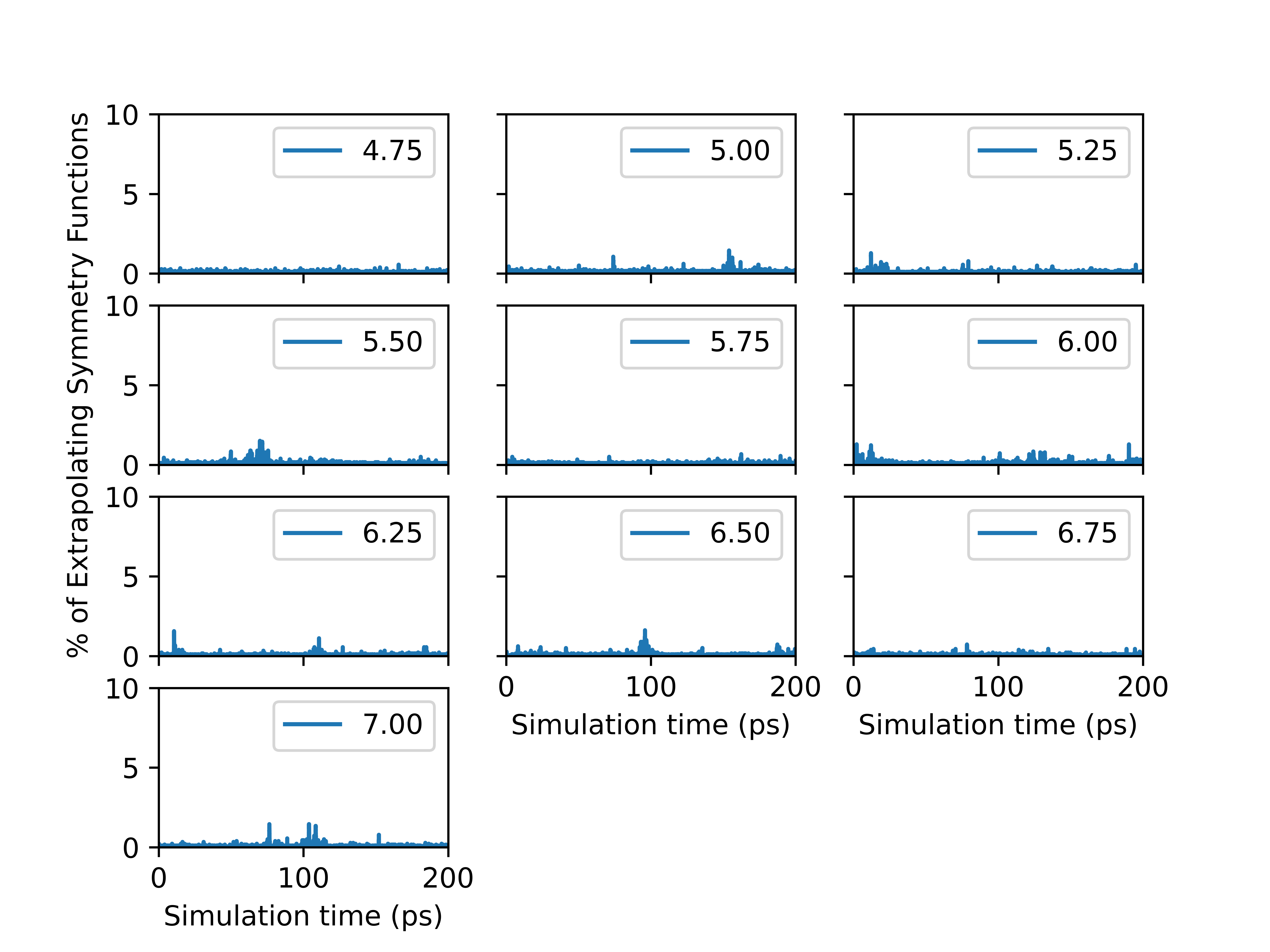}
    \caption{Extrapolation warnings for each window (label corresponds to the restrained distance) from the DNNP part of the MTS-NNP US simulation.}
    \label{ew-baselined-direct}
\end{figure*}
\begin{figure*}[!htp]
    \centering
    \includegraphics[scale=0.8]{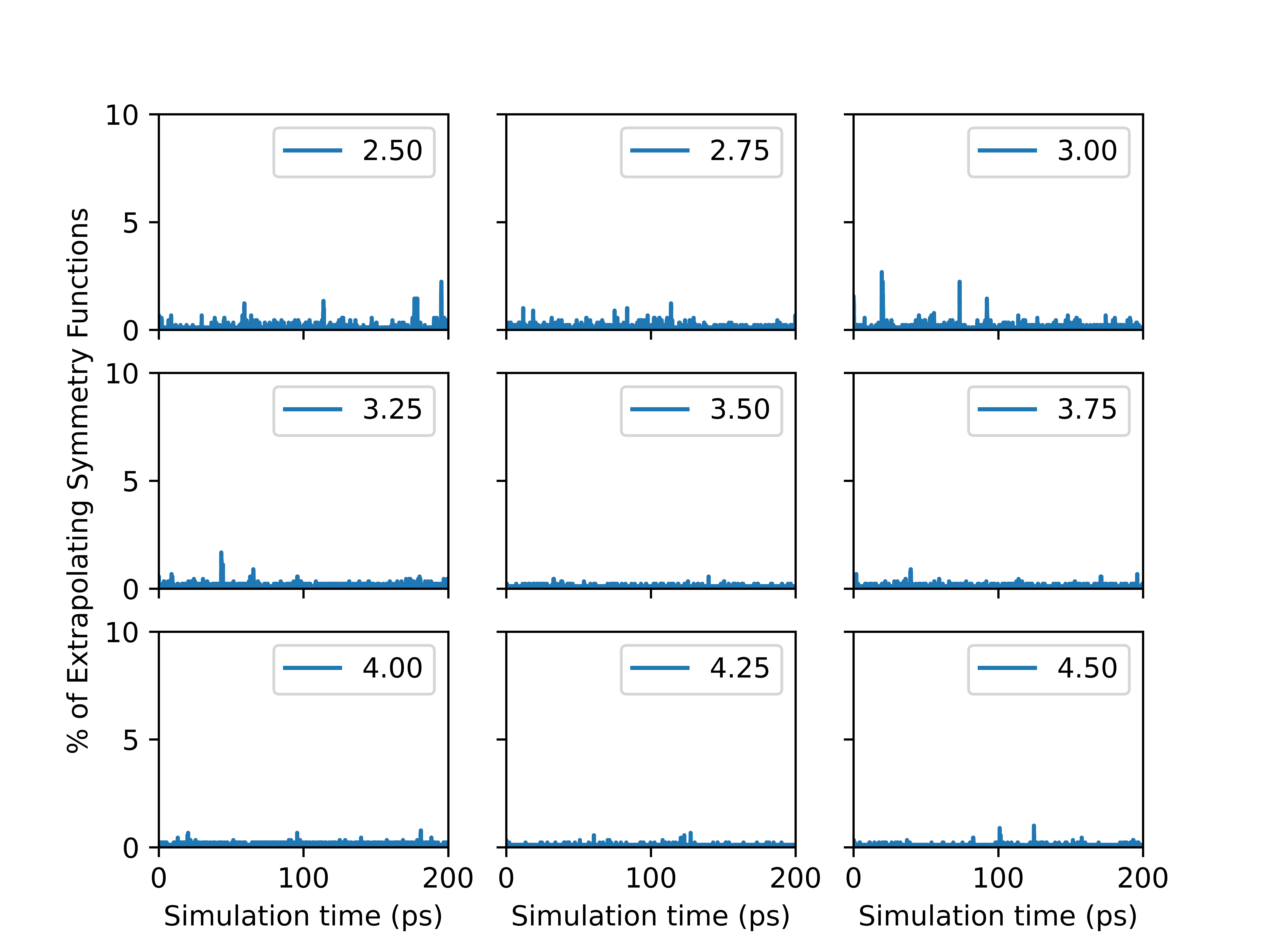}
    \includegraphics[scale=0.8]{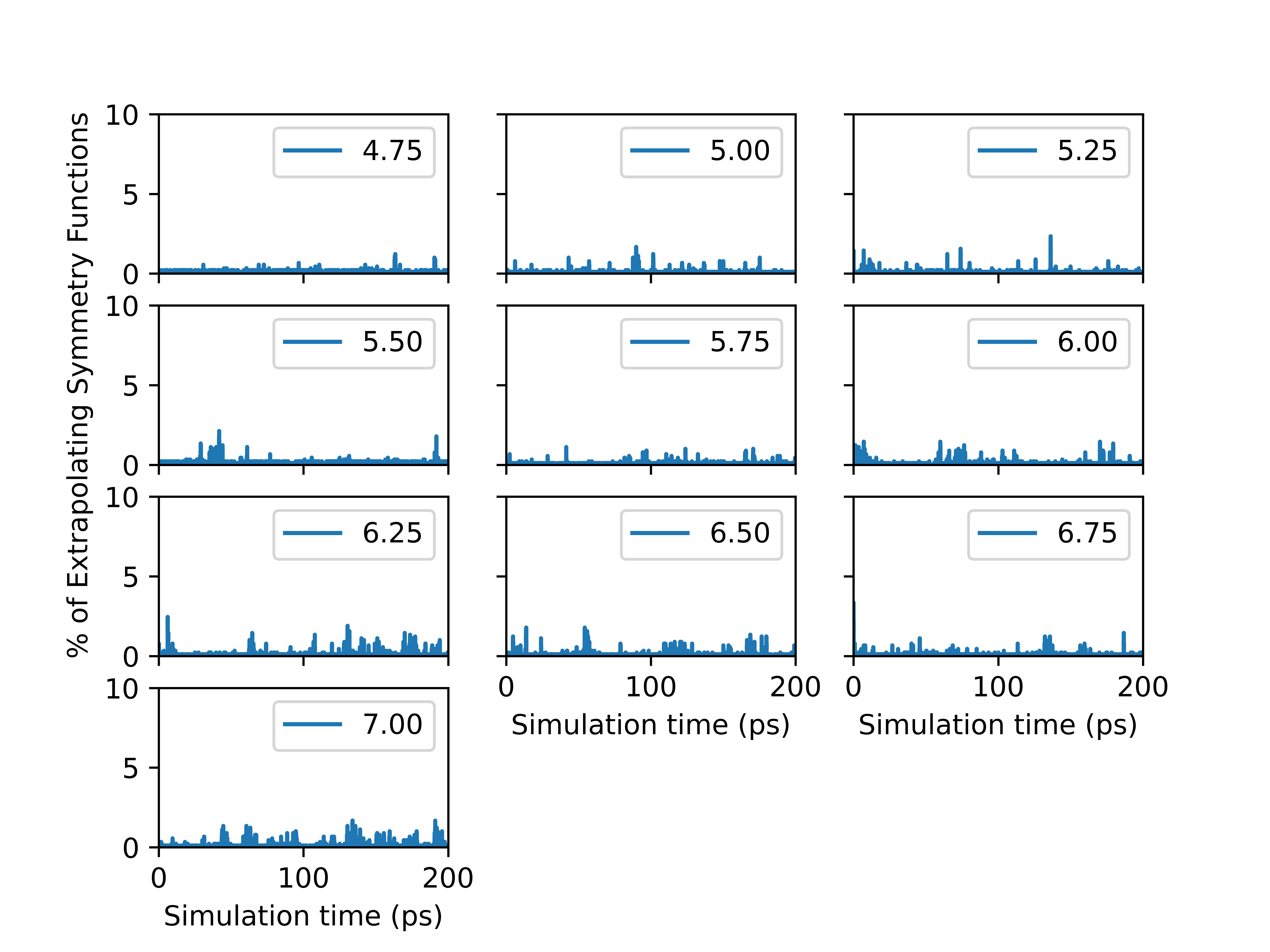}
    \caption{Extrapolation warnings for each window (label corresponds to the restrained distance) from the BNNP part of the MTS-NNP US simulation.}
    \label{ew-baselined-delta}
\end{figure*}
\clearpage

\newpage
\subsection{N-Te-O angle evolution}
\begin{figure*}[!htp]
    \centering
     \includegraphics[scale=0.52]{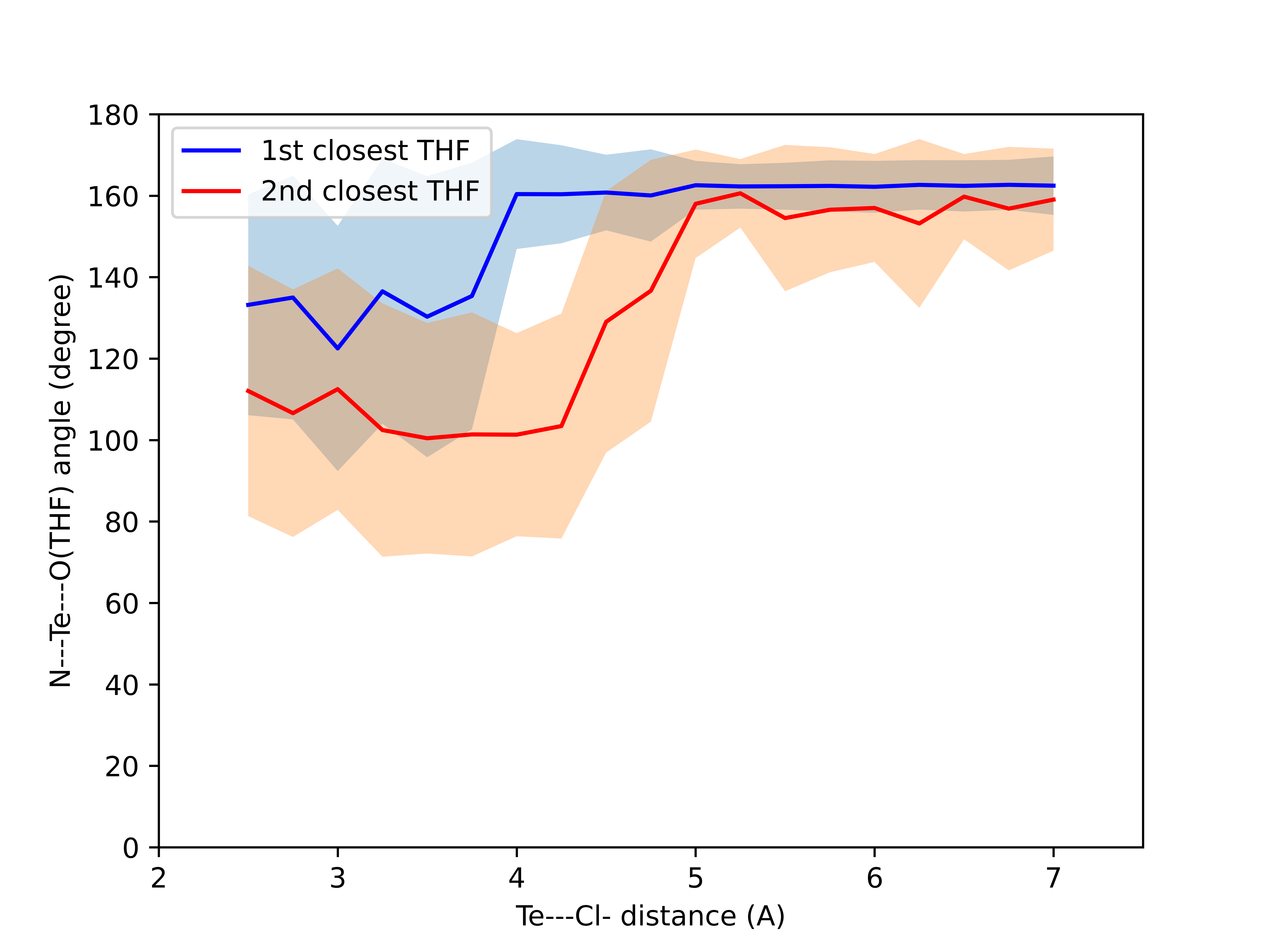}
     \includegraphics[scale=0.52]{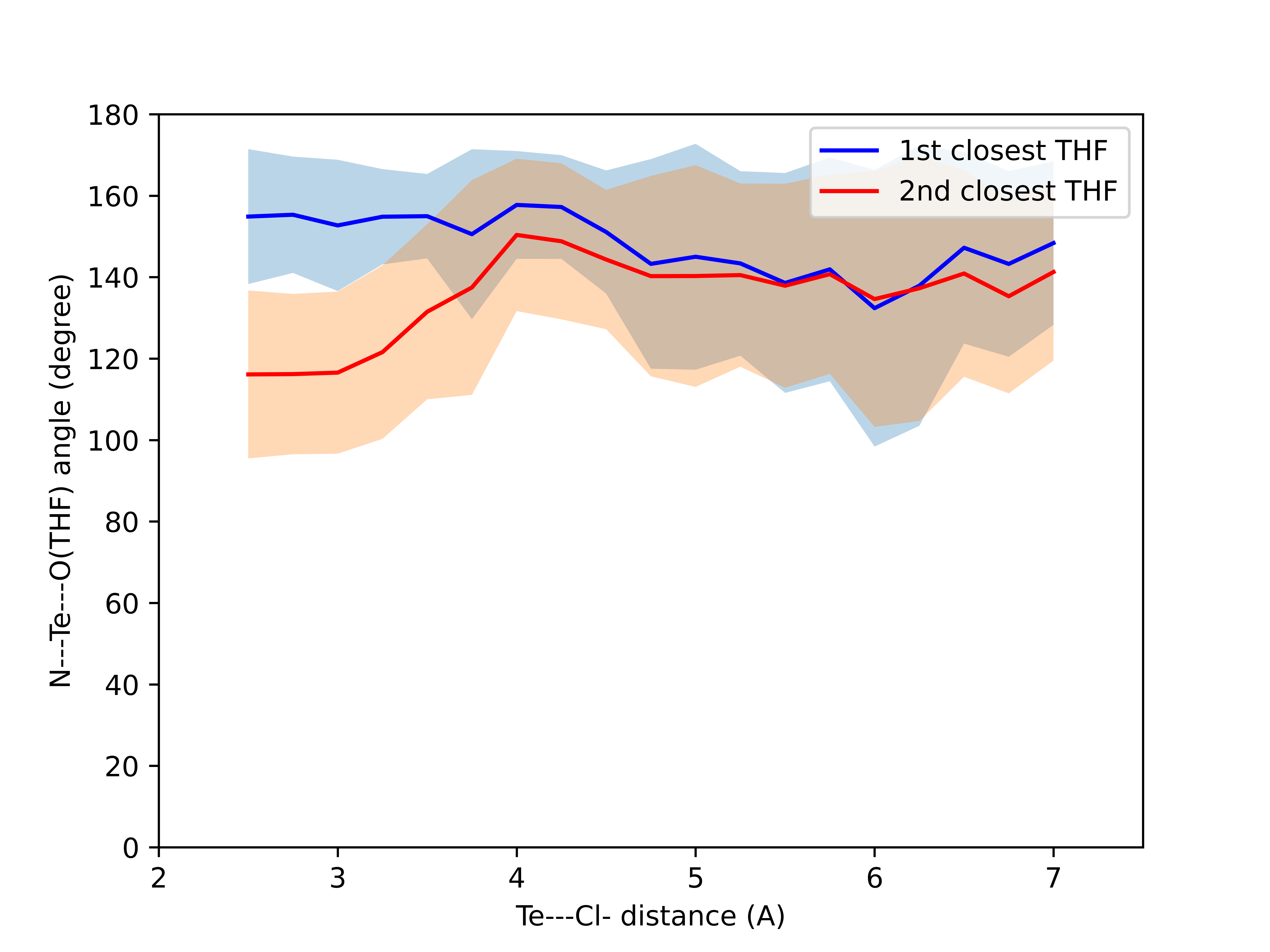}
     \includegraphics[scale=0.52]{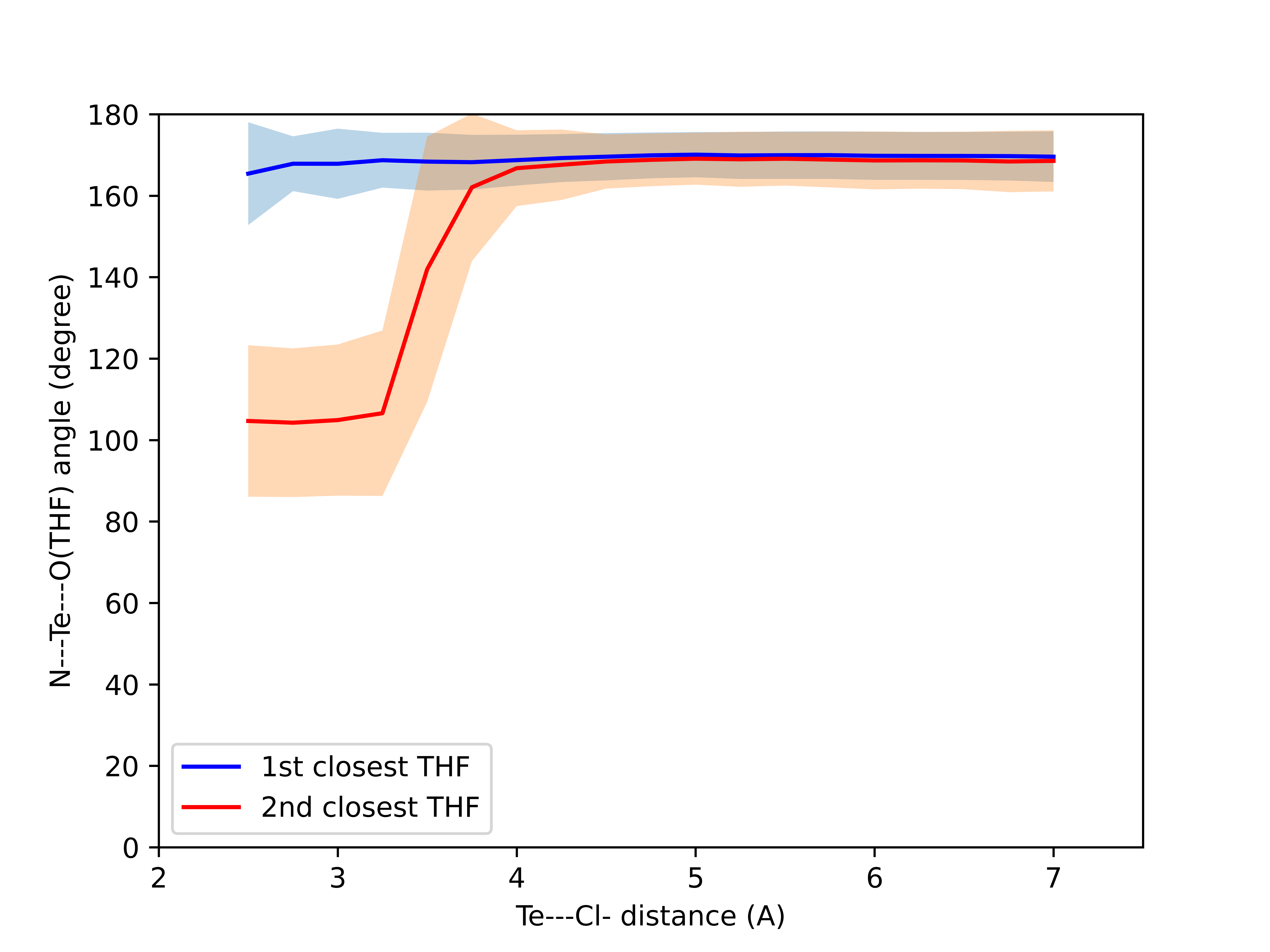}
    \caption{Evolution of the N--Te--O angles for MTS-NNP (top), DNNP (middle) and AMOEBA (bottom) simulations. Atom O belongs to the two THF molecules closest to Te atom. Angle is computed long the Te-\ce{Cl-} distance for every window separately, with standard deviation represented with the shaded regions.}
    \label{angle_evolution}
\end{figure*}

%%%%%%%%%%%%%%%%%%%%%%%%%%%%%%%%%%%%%%%%%%%%%%%%%%%%%%%%%%%%%%%%%%%%%%%%%%%%%%%
\newpage
\section*{References}
%%%%%%%%%%%%%%%%%%%%%%%%%%%%%%%%%%%%%%%%%%%%%%%%%%%%%%%%%%%%%%%%%%%%%%%%%%%%%%%
\bibliography{si}